\documentclass[pdflatex,sn-nature]{sn-jnl}

\usepackage{graphicx}%
\usepackage{multirow}%
\usepackage{amsmath,amssymb,amsfonts}%
\usepackage{amsthm}%
\usepackage{mathrsfs}%
\usepackage[title]{appendix}%
\usepackage{xcolor}%
\usepackage{textcomp}%
\usepackage{manyfoot}%
\usepackage{booktabs}%
\usepackage{algorithm}%
\usepackage{algorithmicx}%
\usepackage{algpseudocode}%
\usepackage{listings}%

\theoremstyle{thmstyleone}%
\theoremstyle{thmstyletwo}%

\theoremstyle{thmstylethree}%

\begin{document}

\title[Article Title]{AI-driven formative assessment and adaptive learning in data-science education: Evaluating an LLM-powered virtual teaching assistant}


\author*[1]{\fnm{Fadjimata} \sur{I.Anaroua}}\email{issoufof@my.erau.edu}

\author[2]{\fnm{Qing} \sur{Li}}\email{qingli2@my.unt.edu}
\equalcont{These authors contributed equally to this work.}

\author[3]{\fnm{Yan} \sur{Tang}}\email{tangy1@erau.edu}
\equalcont{These authors contributed equally to this work.}

\author[4]{\fnm{Hong} \sur{P.Liu}}\email{liuho@erau.edu}
\equalcont{These authors contributed equally to this work.}

\affil*[1]{\orgdiv{Department of Electrical Engineering \& Computer Science}, \orgname{ERAU}, \orgaddress{\street{1 Aerospace Blvd}, \city{Daytona Beach}, \postcode{32114}, \state{FL}, \country{US}}}

\affil[2]{\orgdiv{School of Foreign Languages}, \orgname{Chongqing University of Science and Technology}, \orgaddress{\street{20 East University Town Road}, \city{Shapingba District}, \postcode{401331}, \state{Chongqing}, \country{China}}}

\affil[3]{\orgdiv{Department of Mechanical Engineering}, \orgname{ERAU}, \orgaddress{\street{1 Aerospace Blvd}, \city{Daytona Beach}, \postcode{32114}, \state{FL}, \country{US}}}

\affil[4]{\orgdiv{Department of Mathematics}, \orgname{ERAU}, \orgaddress{\street{1 Aerospace Blvd}, \city{Daytona Beach}, \postcode{32114}, \state{FL}, \country{US}}}


\abstract{This paper presents VITA (Virtual Teaching Assistants), an adaptive distributed learning (ADL) platform that embeds a large language model (LLM)-powered chatbot (BotCaptain) to provide dialogic support, interoperable analytics, and integrity-aware assessment for workforce preparation in data science. The platform couples context-aware conversational tutoring with formative-assessment patterns designed to promote reflective reasoning. The paper describes an end-to-end data pipeline that transforms chat logs into Experience API (xAPI) statements, instructor dashboards that surface outliers for just-in-time intervention, and an adaptive pathway engine that routes learners among progression, reinforcement, and remediation content. The paper also benchmarks VITA conceptually against emerging tutoring architectures, including retrieval-augmented generation (RAG)–based assistants and Learning Tools Interoperability (LTI)–integrated hubs, highlighting trade-offs among content grounding, interoperability, and deployment complexity. Contributions include a reusable architecture for interoperable conversational analytics, a catalog of patterns for integrity-preserving formative assessment, and a practical blueprint for integrating adaptive pathways into data-science courses. The paper concludes with implementation lessons and a roadmap (RAG integration, hallucination mitigation, and LTI 1.3/OpenID Connect) to guide multi-course evaluations and broader adoption. In light of growing demand and scalability constraints in traditional instruction, the approach illustrates how conversational AI can support engagement, timely feedback, and personalized learning at scale. Future work will refine the platform’s adaptive intelligence and examine applicability across varied educational settings.}

\keywords{Distributed Learning, Adaptive Distributed Learning (ADL), Large Language Models (LLMs), ChatGPT, Virtual Teaching Assistants (VITA), GIFT (Generalized Intelligent Framework for Tutoring), Formative Learning Assessment.}



\maketitle

\section{Introduction: The Imperative for Personalized and Scalable Data Science Education}\label{sec1}

The rapid proliferation of data-driven decision making across industries has catalyzed an urgent demand for qualified data scientists, creating unprecedented educational challenges. Educational institutions face mounting pressure to scale their data science programs while maintaining pedagogical quality and responding to the diverse needs of learners from various academic backgrounds. 

This paper presents an innovative approach that addresses this complex educational challenge through integration of OpenAI's large language models with adaptive distributed learning technologies to create personalized, scalable educational experiences. In this introduction, we present the motivation behind our work, provide relevant background information, review pertinent literature, and outline the organization of the paper. 

\subsection{Motivation }\label{subsec1}
The demand for upskilling in data science (DS) far surpasses the capacity of universities, particularly as DS attracts students from diverse undergraduate backgrounds (NIST 2015 and NASEM 2018). While previous efforts have successfully delivered personalized DS education through collaborative learning in small classes, scaling up to meet growing demand presents significant challenges. This paper explores the use of adaptive distributed learning (ADL) and AI-based educational technologies to improve teaching efficiency and scalability without compromising learning outcomes. 

The COVID-19 pandemic accelerated the shift toward distributed learning, blending in-person and online lectures with computer-supported self-study. However, current distributed learning systems exhibit significant limitations in providing the personalized guidance and formative feedback that human instructors deliver in small classroom settings. Standard learning management systems lack the adaptive capabilities necessary for tailoring content to individual learners, while existing automated teaching assistants typically provide generic responses without the contextual awareness needed for deeper learning. To address these scalability challenges, we developed Virtual Teaching Assistants (VITA), leveraging AI tools like OpenAI and ADL technologies developed by the U.S. Department of Défense (DoD). VITA automates formative assessments, provides timely feedback, and supports large-scale hybrid DS courses. Initial simulations demonstrate that VITA can enhance faculty productivity while maintaining educational quality. 

While AI tools like ChatGPT offer solutions to educational scaling, they simultaneously create new challenges to academic integrity (Brenneman and Liu 2024), particularly in distinguishing student-generated work from AI-produced content. Effective assessment design is critical to addressing these challenges while leveraging AI to automate routine teaching tasks and provide online tutoring support. 

\subsection{Background}\label{subsec2}
The curriculum, instructional designs, and preliminary work of the distributed learning environment - iCycle, (intelligent Computer-supported hybrid collaborative learning environment), and early version of BotCaptain have been reported in previous studies (Authors et al 2018, Authors et al 2020, and Authors 2023). The cloud-based courseware and its supported web platform (http://iCycle.cloud) were first deployed in summer 2023. Since then, the ADLT platform has supported courses in Data Visualization and Data Mining, serving students from various institutions in flexible formats (Authors et al 2024). The key innovations of this work include the integration of VITA, automated assessments, visual analytics dashboards, and adaptive learning paths tailored to individual progress. As described in section 3, OpenAI and ChatGPT applications were introduced to the VITA for the first time. Compared to the previous in-house developed BotCaptain, this new version empowered by OpenAI technologies fundamentally transformed the functionalities and user experience.  

The VITA learning environment is built on the Advanced Distributed Learning (ADL) technologies (ADL 2012, and Burmester, 2020) from the U.S. DoD, including data interoperable technology xAPI and GIFT platform for personalized learning and assessment (ADL 2018, Sottilare, Brawner, Goldberg, \& Holden, 2012, Sottilare et al., 2012, Sottilare 2018, Authors 2024).

VITA leverages AI, big data, and distributed learning technologies to offer personalized, anytime-anywhere learning experiences. By utilizing interoperable xAPI (eXperience Application Programming Interface) data standard (IEEE 9274.1.1-2023) and the GIFT framework (Sottilare et al., 2012, Sottilare et al., 2018, and Burmester 2020), the platform tracks student activities, assesses performance, and adapts learning paths. These tools empower instructors to address learning gaps, foster engagement, and enhance the scalability of DS education. This work underscores the transformative potential of combining AI-driven tools and distributed learning technologies for meeting the evolving demands of modern education.

\subsection{Paper Structure}\label{subsec3}

This paper is organized into seven sections. Section 1 introduces the motivation of this study. Section 2 outlines the theoretical grounding of this work - the foundational principles of AI-Enhanced adaptive learning. Section 3 details the new features of the virtual teaching assistant, BotCaptain, powered by OpenAI. Section 4 explains the retrieval and integration of BotCaptain conversation data into the database for learning assessments. Section 5 discusses the use of dashboards to visualize learning activities and implement automated formative assessments. Immediate feedback and adaptive learning experiences for personalized education based on assessments are also presented in Section 5. Section 6 highlights personalized content recommendations and adaptive learning features. Finally, Section 7 concludes the paper by addressing its limitations and outlining future work. 

\section{Foundational Principles of AI-Enhanced Adaptive Learning }\label{sec2}

Our approach to developing VITA is situated at the intersection of multiple theoretical domains that collectively inform its design, implementation, and evaluation. This integrative approach draws from established learning theories while incorporating emerging research on AI-enhanced educational experiences.

\subsection{Theoretical Foundations for Personalized Learning}\label{subsec21}
The quest to provide individually tailored educational experiences has evolved significantly over decades of research and practice. Early educational models operated under the assumption that standardized instruction delivered uniformly would yield consistent outcomes across learners. However, Bloom's (1984) seminal work on the “2 sigma problem" revealed a stark reality: personalized one-on-one tutoring could help the average student perform two standard deviations better than peers in traditional classrooms. This finding highlighted both an opportunity and a challenge—how might we achieve the benefits of personalized instruction at scale? 

Vygotsky's (1978) sociocultural learning theory provided a crucial theoretical foundation with the concept of the zone of proximal development—the gap between what learners can accomplish independently versus with guidance. In traditional settings, human instructors serve as the “more knowledgeable other”, continuously calibrating their support to each student's evolving capabilities. Our work reconceptualizes this dynamic by positioning AI as an adaptive guide that can continuously assess and respond to individual learner needs. 

The practical implementation of personalization principles advanced through Tomlinson's (2001) differentiated instruction framework, which articulated how educational experiences could be thoughtfully varied across content, process, and products based on each learner's readiness, interests, and learning profile. While originally developed for human teachers, these principles offer a blueprint for what AI-powered systems might achieve with greater precision and at unprecedented scale. 

The importance of dialogue in learning finds theoretical expression in Laurillard's (2013) Conversational Framework, which conceptualizes learning as an iterative dialogue between teacher and student. This framework illuminates how understanding develops through cycles of discursive interaction, adaptation, reflection, and reconceptualization—processes that our BotCaptain system aims to emulate through AI-powered Socratic questioning. 

\subsection{From Theory to Technology: Adaptive Learning Systems}\label{subsec22}
The challenge of translating personalization principles into technological systems has evolved through increasingly sophisticated approaches. Early adaptive systems like Corbett and Anderson's (1995) ACT Programming Tutor demonstrated how computer-based instruction could adapt to individual student models, but these systems were constrained by predetermined pathways and limited assessment capabilities. 

Addressing some of these limitations, Vandewaetere et al. (2011) later developed a comprehensive model of adaptive e-learning, conceptualizing adaptation across learner characteristics, activities, outcomes, and contexts. Their work highlighted a critical tension that remains central to our research: balancing system-controlled adaptation with learner agency in the personalization process. 

FitzGerald et al. (2018) further expanded this understanding by identifying critical dimensions of personalization: what is being personalized (content, pace, assessment), who controls the personalization (system, teacher, learner), and the purpose of personalization (efficiency, effectiveness, engagement). Their research revealed that previous adaptive systems often excelled in one dimension while neglecting others—adapting content effectively but surrendering control entirely to the system, or preserving learner agency but compromising on personalization accuracy. 

A promising resolution to this tension emerges from the integration of Self-Regulated Learning theory (Zimmerman \& Schunk, 2011) with adaptive technologies. Chen et al. (2021) demonstrated that effective personalized systems not only adapt content based on performance but also develop learners' capacity to monitor and regulate their own learning processes. This insight directly informs our implementation of VITA, which combines system-driven adaptivity with features designed to promote metacognitive awareness and self-direction. 

\subsection{Ethical Considerations in AI-Enhanced Learning Environments }\label{subsec23}

The implementation of AI-powered learning systems introduces complex ethical considerations that require careful attention. As we develop systems that collect extensive learner data to enable personalization, questions of privacy, autonomy, and equity become increasingly significant. 

Prinsloo and Slade's (2017) ethical framework for learning analytics provides a critical foundation, emphasizing that data collection must be balanced with student agency and transparency. Their work challenges the assumption that more data automatically leads to better personalization, arguing instead for purposeful collection aligned with clearly articulated educational goals. This perspective has directly informed our approach to xAPI data collection, which prioritizes meaningful learning events over indiscriminate tracking. 

Academic integrity in AI-enhanced environments presents particularly complex challenges. Bretag et al.’s (2020) framework for addressing contract cheating remains relevant, but the emergence of generative AI tools has introduced new vulnerabilities and blurred traditional lines of authorship. Contemporary approaches to this challenge emphasize authentic assessment, promoting approaches that maintain validity even when students have access to sophisticated AI tools. Our implementation addresses these concerns through authentic assessment, iterative feedback processes, and tasks that emphasize critical thinking over reproduction. 

Our implementation navigates a particularly complex ethical terrain: we simultaneously leverage AI's capabilities for educational benefit while implementing safeguards against potential misuse. This duality is reflected in our system design, which incorporates Tsai et al.'s (2020) principles for ethical AI in education—emphasizing transparency, maintaining human oversight of AI-driven decisions, and ensuring that personalization serves educational goals rather than technological convenience. 

\subsection{Learning Analytics for Personalized Educational Experiences }\label{subsec24}

The ability to collect, analyze, and act upon learner data represents the technical foundation of effective personalization. Our implementation of xAPI-based learning analytics builds upon theoretical work by Siemens (2013), who emphasized that effective analytics requires not just data collection, but purposeful analysis oriented toward improving learning outcomes. 

Ferguson's (2012) comprehensive review of the learning analytics field highlighted the disconnect between technological capabilities and pedagogical applications—a gap our integration of OpenAI with xAPI aims to bridge. By analyzing not just correctness but the quality and progression of dialogue, our system develops a richer understanding of learner needs than would be possible through conventional assessment alone. 

The xAPI standard itself represents a significant advancement in how learning data is conceptualized and structured to support data exchange across different online learning platforms. Our implementation follows Bienkowski et al.'s (2012) guidelines for educational data interoperability while addressing their concern that data collection must be aligned with clear learning objectives. This alignment ensures that adaptation decisions are based on pedagogically relevant information rather than incidental interactions. 

Lang et al. (2017) identified several critical measurement challenges in learning at scale, particularly regarding the interpretation of complex learning behaviors. Their work highlights how traditional assessment metrics often fail to capture the nuanced progression of understanding—a limitation we address through our integration  

This paper is intended primarily for scholars in learning science, data science, and educational technology. As such, we provide only a brief overview of foundational technologies like ADL and GIFT, focusing instead on their role in enabling adaptive learning. Detailed technical configurations and system settings are available in our Gitbook for interested readers of conversational AI with xAPI tracking. 

\subsection{Theoretical Intersections and Unexplored Terrain}\label{subsec25}
Our review of the literature reveals four critical gaps that existing research has not adequately addressed:(1) the absence of practical implementations that leverage advanced large language models (LLM) to realize established personalization theories; (2) the disconnect between adaptive learning systems and natural language capabilities needed for effective dialogic instruction; (3) underdeveloped ethical frameworks for maintaining academic integrity in AI-enhanced environments; and (4) limited learning analytics models for deriving personalization insights from conversational data. 

VITA addresses these gaps by integrating OpenAI's language capabilities with established learning frameworks, creating an implementation that advances both theoretical understanding and practical applications of AI-enhanced personalized learning. By building upon established educational theory while addressing these underexplored areas, our work contributes to a more comprehensive understanding of how conversational AI can transform educational practice in data science education. 

\section{OpenAI-enhanced BotCaptain - a VITA Redesigned for Personalized Instruction}\label{sec3}

This section outlines the integration of the OpenAI Chat Block (BotCaptain) into our interactive learning platform Moodle, emphasizing its customization and pedagogical applications in data science education. We highlight BotCaptain’s core capabilities as a VITA, including the use of tailored prompts for tasks such as summarization, explanation, comparison, critical analysis, role-playing, and problem-solving. Special attention is given to the Socratic dialogue feature—its design, implementation, and role in fostering critical thinking. We also illustrate how BotCaptain enhances learning through 24/7 availability, context-aware responses, and automated formative feedback.  

\subsection{Interactive Learning Management System Conceptual Model }\label{subsec31}

The conceptual model of the ADL platform is shown in Figure 1. All AI-based learning technologies presented in this paper use xAPI assertions to track and record learning activities (Advanced Distributed Learning 2018, https://xAPI.com, IEEE 9274.1.1-2023). An xAPI assertion, using the JavaScript Object Notation (JSON), can be mapped into a uniform "who did what" format with optional auxiliaries.  For instance, in the xAPI assertion “Susan completed Data Science Ethics (Novice)”, the subject “Susan” is a student, and verb is predefined with more details below.  The underscored object “what” links to the predefined knowledge, skills, task, or content with clear learning objectives. The auxiliary segment is used to determine competency level: Novice, Intermediate, or Expert. As structured data with predefined vocabulary, xAPI assertions are not only interoperable, but can also be quickly processed and “understood” by computers. 

\begin{figure}[h]
\centering
\includegraphics[width=0.9\textwidth]{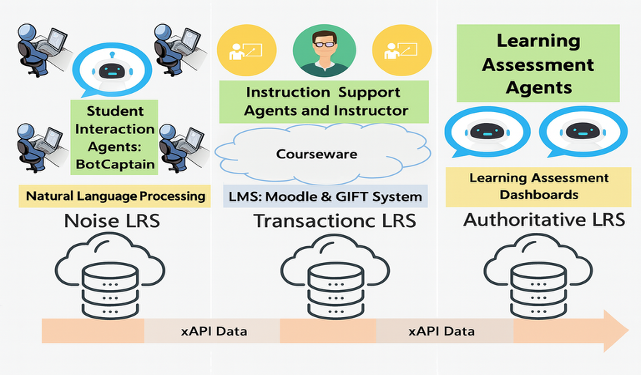}
\caption{Conceptual Model of AI-Augmented ADL Platform}\label{fig1}
\end{figure}

An xAPI assertion supports variable granularities to assess multiple levels of nested competencies that are stored in a chain of Learning Record Stores (LRS) such as Noise LRS, Transactional LRS, and Authoritative LRS. For example, a low-level event such as “Susan viewed a video clip” is stored in a noised LRS of LMS while a highest level of competency such as “Susan asserted DM course (A)” is stored at an authoritative LRS. Such learning activities at all granularities provide robust evidence to assess learning outcomes. 

The System Components and Deployment Architecture is shown in Fig. 2 (Authors 2024). We used Articulate 360 Storyline for content creation, Amazon S3 for content hosting, and GIFT and Moodle for content delivery. Articulate Storyline 360 features branching scenarios for tailored learning experiences (see Figure 15). Content is then uploaded to Amazon S3 for secure, scalable storage. The hosted content is integrated with GIFT, which is deployed on Amazon AWS EC2, and connected to Moodle using Learning Tools Interoperability (LTI) such as xAPI data exchange. This setup allows GIFT to dynamically adjust content based on learner interactions, providing personalized learning experiences while LRS captures detailed activity data. The iCycle platform uses the LRS (https://erau.xapi.io ) donated by Veracity Learning Technologies Consulting LLC. Therefore, we will refer to our LRS hereinafter as Veracity LRS. 

\begin{figure}[h]
\centering
\includegraphics[width=1.0\textwidth]{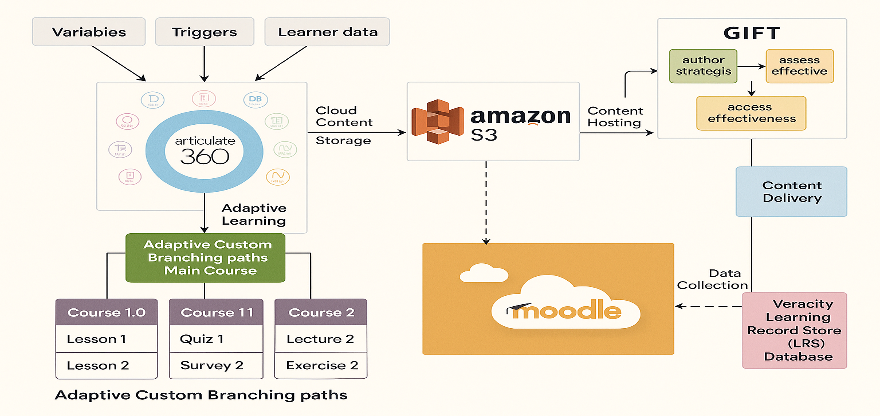}
\caption{The System Components and Deployment Architecture}\label{fig2}
\end{figure}

\subsection{Integration of OpenAI Chat Block (BotCaptain) into Moodle }\label{subsec32}

To enhance our interactive learning, an OpenAI chat block (New features of BotCaptain) with custom completion prompts setting is integrated into the Moodle website, facilitating real-time, AI-driven assistance and tutoring for students. This feature leverages OpenAI’s conversational AI to provide immediate, contextual support on various topics within each Moodle course. The implementation process involved embedding the chat block within Moodle’s framework, with a custom Learning Tools Interoperability (LTI) connection ensuring seamless communication between Moodle and the OpenAI model.  

After initial setup, the chat block was tested with sample questions to verify response accuracy and alignment with course objectives. Adjustments could be made as needed to optimize the model’s responses, ensuring clear, concise, and constructive assistance for students. Detailed technical configurations and system settings are available in our Gitbook for interested readers of conversational AI with xAPI tracking and xAPI to LRS Data Streaming. 

\subsection{Data Retrieval and Integration with LRS}\label{subsec33}
 This section highlights how AI supports formative assessment and provides instructors with real-time insights through learning analytics. Strategies like self-referential, information-starved, and temporally based queries help ensure meaningful student engagement while minimizing misuse of generative AI. Socratic dialogues are also evaluated to promote critical thinking. LRS dashboards visualize learning activity, enabling instructors to track progress, spot outliers, and intervene effectively using data from chat logs and learner interactions. 

The chat block interactions are visualized on the existing LRS learning analytics dashboard, allowing instructors to analyze student engagement patterns and identify common questions or recurring topics of confusion. This data-driven approach enables instructors to proactively address knowledge gaps and provide targeted support, ultimately fostering a more personalized learning experience. This OpenAI chat integration exemplifies an advanced application of AI within an LMS, offering flexible, scalable, and tailored support to enhance student learning outcomes. 
 
\subsection{Open AI/BotCaptain Chat Logs Reports and xAPI}\label{subsec34}
To monitor the chat block’s impact on student learning, xAPI statements were configured to record interactions between students and the OpenAI chat feature. As shown in Fig. 3, each interaction logs the query, the AI’s response, and the timestamp, capturing a detailed history of student engagement and support needs. This data can be sent to the LRS alongside other learning activities, offering instructors additional insights into student understanding and areas requiring further clarification.

\begin{figure}[h]
\centering
\includegraphics[width=1.0\textwidth]{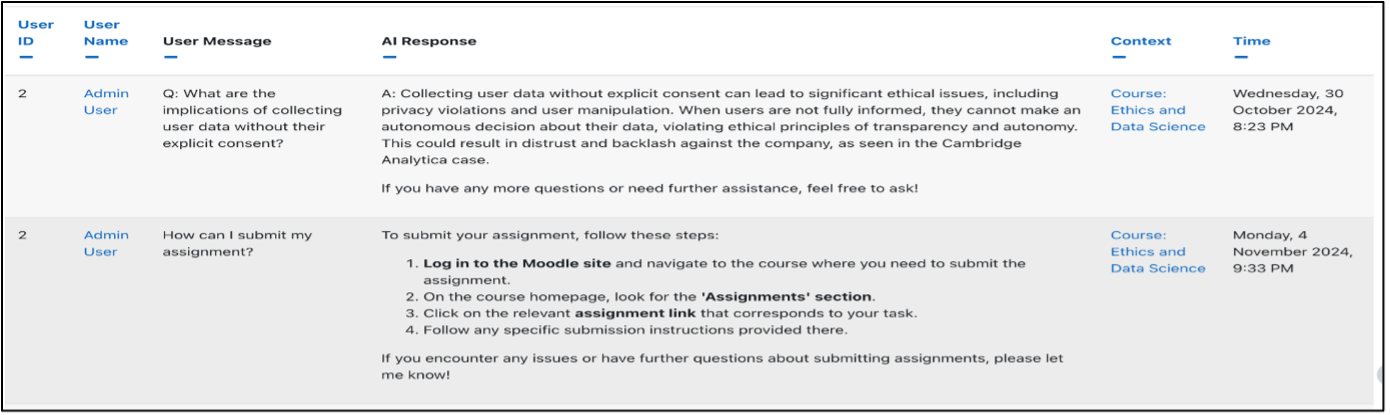}
\caption{Chat Logs Reports Overview}\label{fig3}
\end{figure}

\subsection{Transformation and Upload of OpenAI Chat JSON file to Veracity LRS}\label{subsec331}

This section outlines the streamlined process for integrating OpenAI chat logs into the Veracity LRS. To enable effective learning analytics, a two-step data integration process is implemented to convert chatbot interaction logs (shown in Fig.3) into xAPI statements and upload them to the Veracity LRS (Fig.4). This pipeline ensures structured tracking of learner activity and supports future scalability through reusable scripts, secure credential handling, and error-resilient uploads. By transforming raw JSON data into xAPI-compliant statements and securely uploading them to the LRS, this approach ensures accurate tracking and analysis of user interactions. The integration leverages automated algorithms included in our Gitbook to enhance efficiency and maintain compliance with xAPI standards, enabling comprehensive insights into engagement and learning outcomes. 

\begin{figure}[h]
\centering
\includegraphics[width=1.0\textwidth]{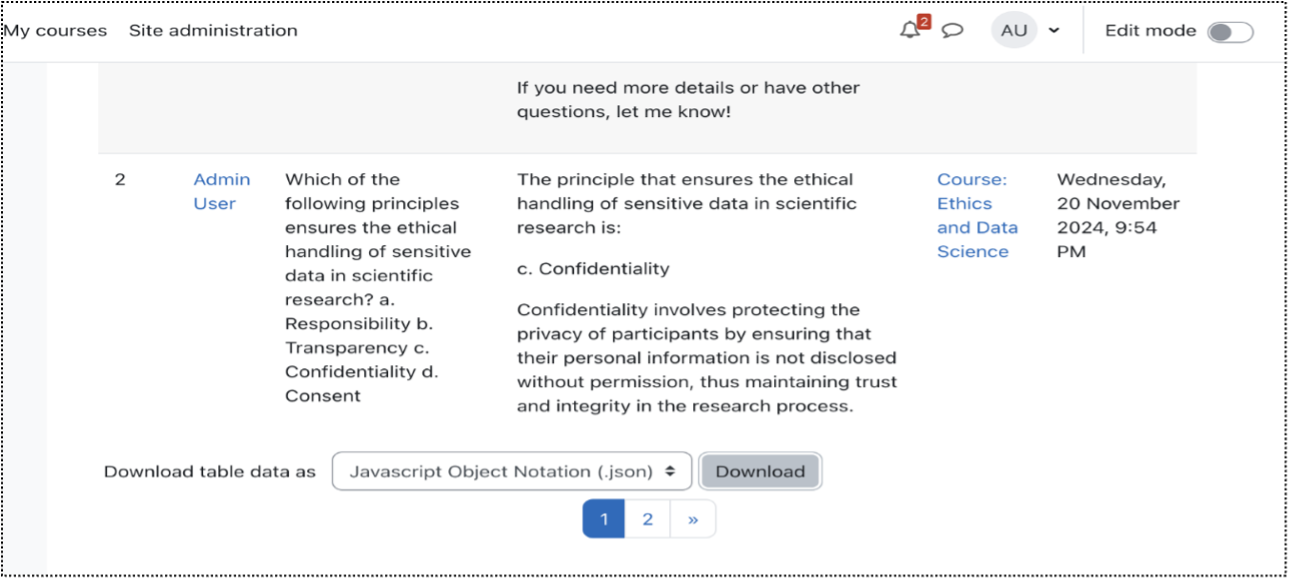}
\caption{OpenAI Chat Log report (JSON File) from Moodle LMS/Moodle Log Report File Structure}\label{fig4}
\end{figure}

The next step is for transforming JSON to xAPI structure and checking Statements. The Fig. 4 displays a sample xAPI statement in JSON format, representing a logged learner interaction within a Moodle-based chatbot environment. It includes key components such as the actor ("Admin User"), the verb ("asked"), and the object (a question on data privacy). The statement is timestamped and structured to meet xAPI specifications, enabling it to be transmitted to a Learning Record Store (LRS) for tracking and analysis.

Finally, the Login Automation and xAPI Statements data feed to Veracity LRS (shown in Fig. 5) will support meaningful analysis of student engagement, where chatbot interaction data is transformed into xAPI statements and transmitted to the LRS. 

\begin{figure}[h]
\centering
\includegraphics[width=1.0\textwidth]{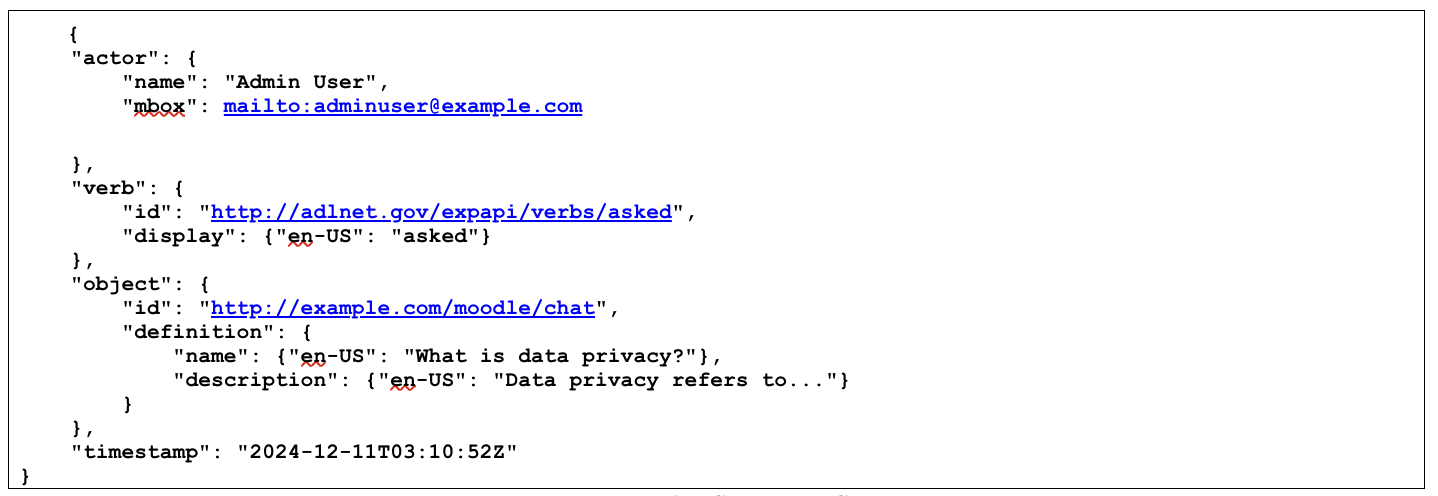}
\caption{Example of xAPI Statements Structure}\label{fig5}
\end{figure}

The process displayed in Fig. 6 enables fine-grained tracking of AI-supported learning interactions, offering insights into learner behavior, dialogue quality, and system effectiveness. The workflow was designed for scalability and reuse, with secure credential handling and structured validation to ensure data integrity and reliability. 

\begin{figure}[h]
\centering
\includegraphics[width=1.0\textwidth]{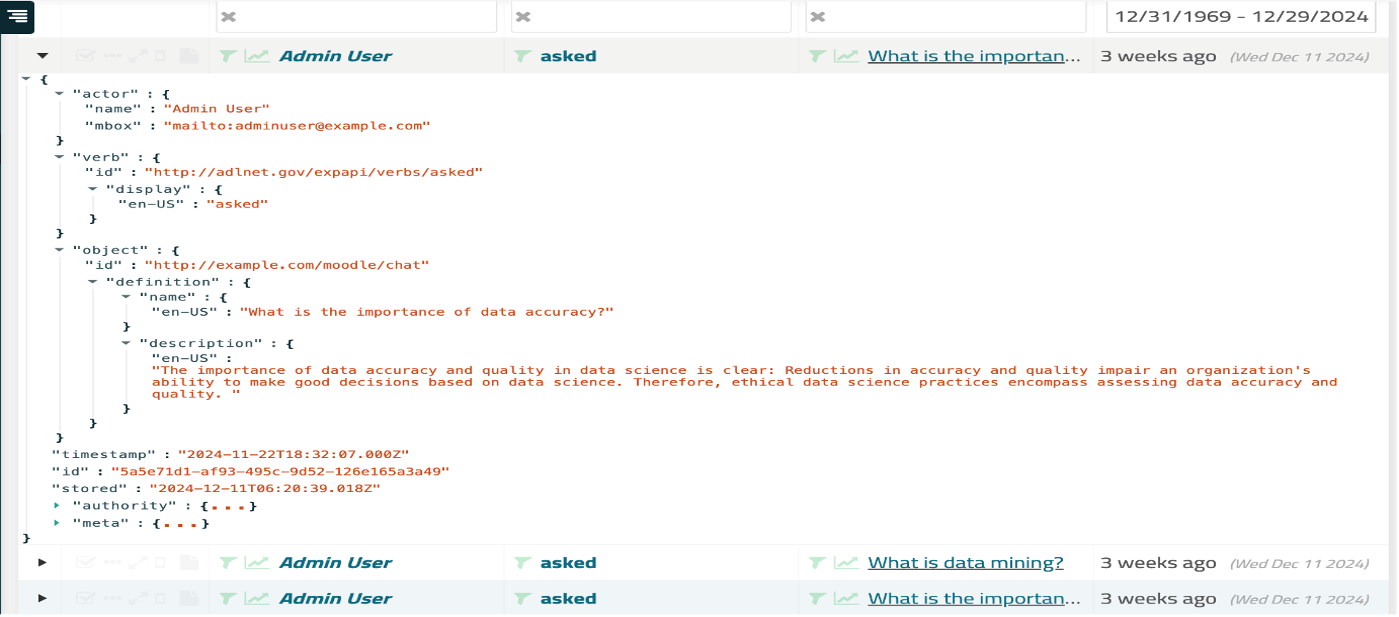}
\caption{Login Automation and xAPI Statements data feed to Veracity LRS}\label{fig6}
\end{figure}

Using Moodle’s analytics tools, instructors can monitor students' progress in understanding core ethical principles and adjust the tutoring focus as necessary. Data from quiz performance and student interactions during tutoring can be tracked to provide a holistic view of student comprehension. By blending Socratic questioning with structured tutoring sessions focused on the lessons, we can deepen students' understanding of the course module while also creating an interactive learning environment that promotes critical thinking. 

\subsubsection{Automated Report Extraction Using Web Scraping}\label{subsubsec31}

Since Moodle's Web Services API has limitations in providing direct access to chat or report data, an automated web scraping approach was implemented using Selenium. This method shown in Fig. 7 simulates user interaction with the Moodle interface to log in, navigate to chat reports, and extract relevant data. The retrieved files are then processed and formatted for transmission to the LRS, enabling integration of learner interaction data despite API limitations. 
\begin{figure}[h]
\centering
\includegraphics[width=0.3\textwidth]{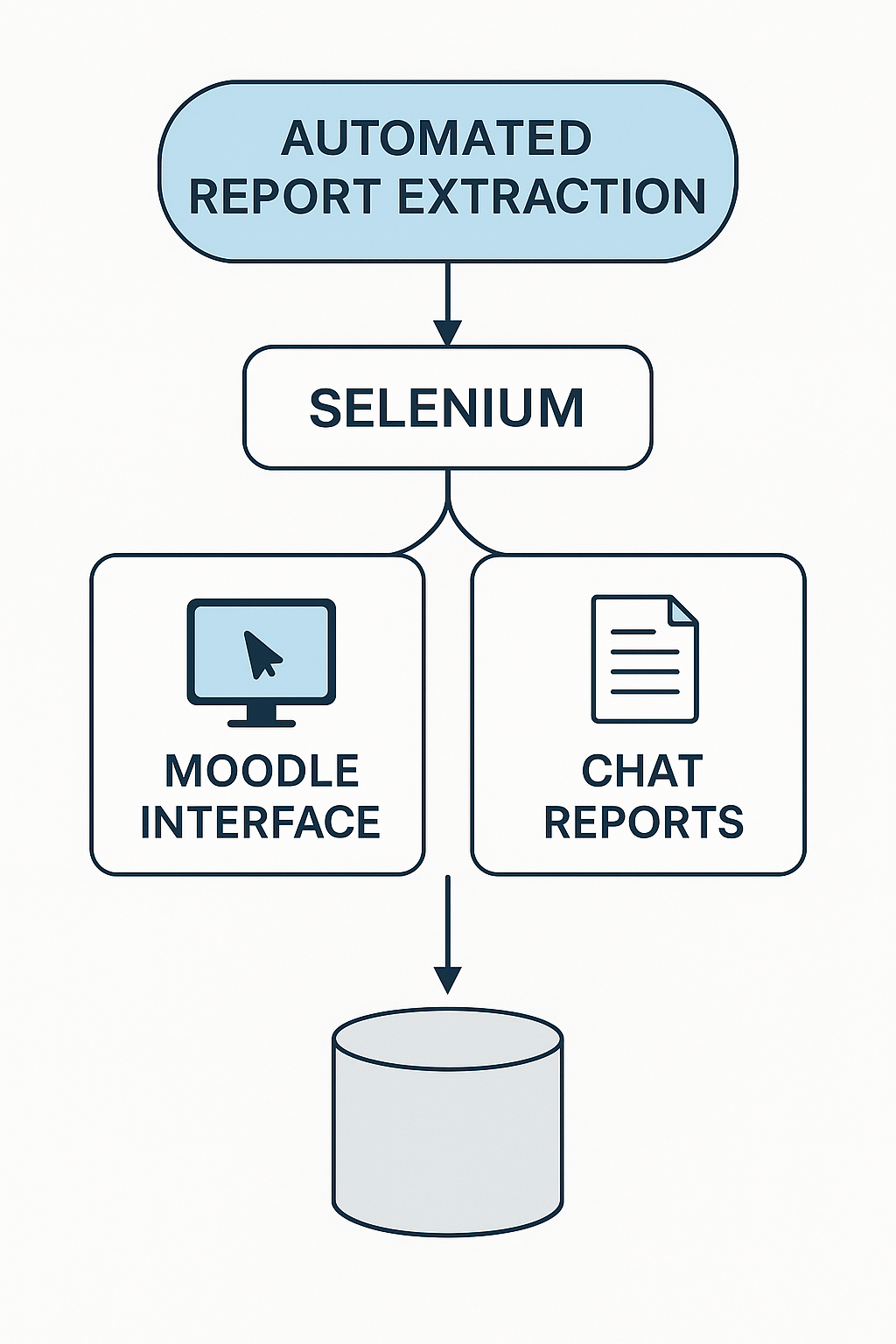}
\caption{Automated Report Extraction from Moodle Using Web Scraping and Selenium}\label{fig7}
\end{figure}

\subsubsection{BotCaptain Pedagogical Benefits}\label{subsubsec32}
The integration of OpenAI’s chat block (BotCaptain) offers significant pedagogical benefits: 
\begin{itemize}
\item\textbf{Availability}: Students can access the chat block at any time, promoting self-paced learning and allowing them to seek help even outside of class hours. This is particularly beneficial for asynchronous courses, where students may have questions after reviewing course materials independently. 

\item\textbf{Context-Aware Responses}: Through machine learning, the chat block provides responses that are contextually relevant to each course module, based on keywords and course-specific terminology. This targeted approach minimizes irrelevant information, guiding students toward resources directly related to their queries and adapting to their level of understanding. 

\item\textbf{Automated Summative Feedback}: The chat block is also equipped to provide automated feedback based on common questions and areas where students typically seek help. Notifies instructors of recurring topics or challenges, allowing for timely adjustments to course materials and instructions. 
\end{itemize}

\section{Implementation of the ChatBot Features of BotCaptain.}\label{sec4}

To effectively personalize instruction and engage learners in data science education, this section focuses on BotCaptain features that employ tailored prompts aligned with specific pedagogical objectives. These prompts form the foundation of contextualized AI interactions that support comprehension, critical thinking, and active learning. By framing student-AI conversations with well-defined educational intents such as summarizing content, practicing concepts, or exploring ethical scenarios, the system ensures meaningful engagement.

\subsection{Student Support with Customised Prompts Dialogue}\label{subsec4}

This section explores the strategies used to structure and implement these prompts, highlighting how prompt engineering enhances the instructional value of BotCaptain within adaptive learning environments. 

\subsubsection{Defining the Purpose of the Open AI Block Completion Prompt}\label{subsubsec4}

As described in Table 1, for educational goals, it is important to clarify the purpose of the interaction. The system determines whether we are helping students understand a concept, generate ideas, summarize content, or practice skills. To match the audience level, we tailor the prompt to our students' knowledge level. For example, a prompt for beginners might use simpler language, while a prompt for advanced learners can be more complex and specific. 

\begin{table}[h!]
\centering
\caption{Prompt Scenarios and Dialogue Specifications for Instructional Interactions}
\label{tab:prompt_scenarios}
\begin{tabular}{@{}p{4cm} p{8.5cm}@{}}
\toprule
\textbf{\textit{Prompt Scenarios}} & \textbf{\textit{Dialogues Specification}} \\
\midrule

\textbf{A. Summarization} & \parbox[t]{8.5cm}{“Summarizing the key points of [Topic/Article] \\
in a concise paragraph.”} \\ \\

\textbf{B. Explanation} & \parbox[t]{8.5cm}{“Explaining [Concept] as if teaching to someone \\
with no prior knowledge. Using simple language and \\
providing examples.”} \\ \\

\textbf{C. Comparison} & \parbox[t]{8.5cm}{“Comparing and contrasting [Concept A] and [Concept B]. \\
Listing three main similarities and three differences.”} \\ \\

\textbf{D. Critical Analysis} & \parbox[t]{8.5cm}{“Evaluating the effectiveness of [Theory or Method] \\
and discussing its strengths and weaknesses.”} \\ \\

\textbf{E. Role-Playing} & \parbox[t]{8.5cm}{“Imagining being [a famous scientist, historical figure, etc.]. \\
Explaining this perspective on [Topic].”} \\\\

\textbf{F. Problem-Solving} & \parbox[t]{8.5cm}{“Given [Scenario], proposing a solution. \\
Justifying choice with at least two reasons.”} \\ \\

\bottomrule
\end{tabular}
\end{table}

\subsubsection{Structure the Prompt for Context and Focus}\label{subsubsec41}

To provide context for more tailored responses, we start with a context or scenario, followed by a specific request. Example: "Imagine explaining the concept of photosynthesis to a middle school student. Summarize the process in simple terms.” 

\subsubsection{Chat Block Completion Prompt \& Source of Truth initial stage Configuration}\label{subsubsec42}

The OpenAI Chat completion prompts, and source of truth will provide tailored, context-specific interactions that help students engage more deeply with course content. When prompts are carefully crafted, students receive clear, concise explanations, summaries, or analyses based on the specified instructions. 

\begin{figure}[h]
\centering
\includegraphics[width=1.0\textwidth]{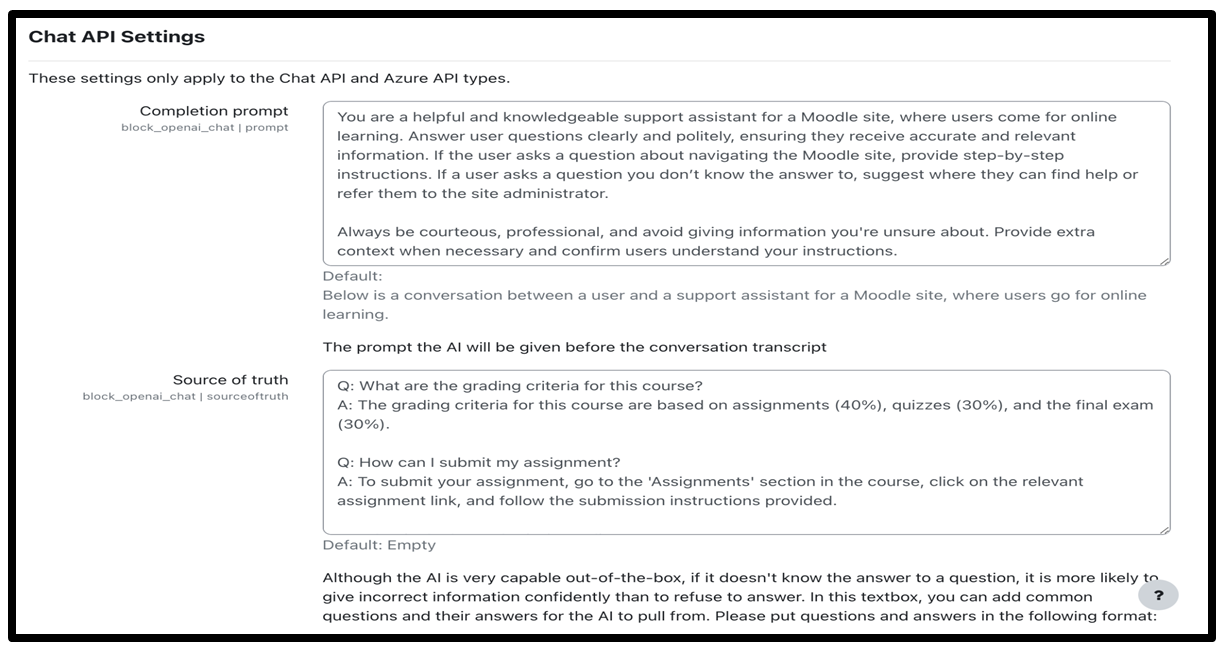}
\caption{BotCaptain AI sample initial prompts for Course Q\&A}\label{fig8}
\end{figure}

\subsubsection{Open AI Chat/BotCaptain responses based on prompts input}\label{subsubsec43}

A well-defined prompt asking for a summary of a course specific event results in a brief, factual paragraph highlighting key points, while a prompt designed to elicit critical thinking encourages students to analyze or evaluate concepts. This guided response system described in Fig. 9 helps students understand complex topics in a more structured manner, making the learning experience interactive. 

\begin{figure}[h]
\centering
\includegraphics[width=1.0\textwidth]{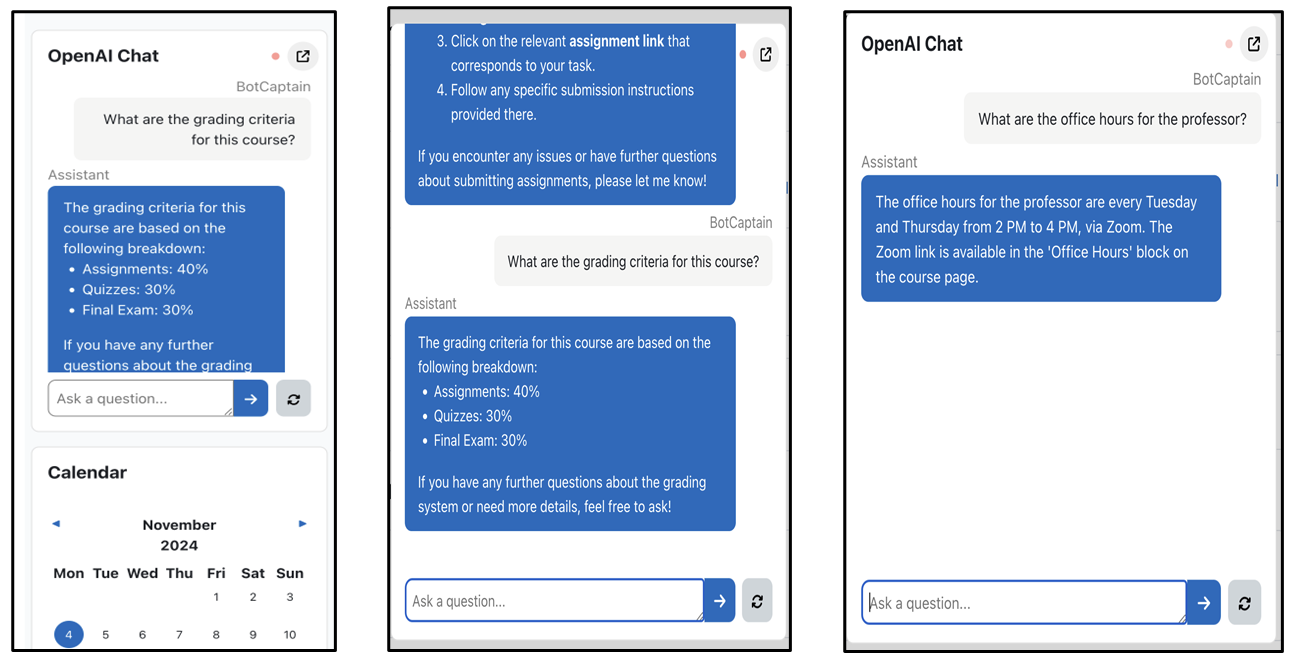}
\caption{Open AI Chat/BotCaptain sample responses from Completion Prompts setting}\label{fig9}
\end{figure}

\subsubsection{Setting up the Stage for Initial/Socratic Prompt Dialogue}\label{subsubsec44}

The next step and goal is to set up Socratic dialogue by encouraging critical thinking and framing questions in a Socratic manner. Instead of providing direct answers, the AI tutor can guide students toward the correct understanding by asking probing questions.

\subsection{Design Process for Interactive Socratic Prompts Dialogue}\label{subsec41}

In this subsection, the completion prompt configuration processes will be redirected to a more customised or Socratic design. Instead of just feeding the BotCaptain AI chat block with direct answers, we intend to create prompts that ask students to reflect on ethical dilemmas or concepts.

\textit{Example 1 Simple Prompts for an Ethics and Data Science Course/Lesson 1} 
\begin{itemize}
    \item \textbf{Prompt 1/ Q1:} "What do you think would happen if a data scientist knowingly discards conflicting data? How might this affect the trust in their research?" 
    \item \textbf{Prompt 2/ Q2:} "Do you believe it is ever acceptable to manipulate data to achieve desired results? Why or why not?" 
    \item \textbf{Prompt 3/ Q3:} "In what ways can integrity guide a data scientist when they face ethical dilemmas? Can you think of real-life examples?" 
\end{itemize}

From these simple questions, we input direct answers in the source of truth of our Chat Block, but for Socratic prompts we follow up with Socratic dialogues tailored to help students deepen their understanding through thoughtful questioning. 

\textit{Example 2 Socratic Dialogue for the Lesson 1 on Confidentiality}
\begin{itemize}
    \item \textbf{Tutor's Question 1:} "How do you think informed consent in data collection influences the ethical landscape of a data science project?" 
    \textit{Follow-up:} "Can you think of any real-world examples where failure to get consent led to ethical concerns?" 
    \item \textbf{Tutor's Question 2:} "What do you think happens when a company fails to maintain confidentiality of user data?" 
    \textit{Follow-up:} "How could this failure affect both the company and the individuals whose data is exposed?" 
\end{itemize}

\subsubsection{Socratic Dialogue Example Scenario Development}\label{subsubsec45}

\textit{Topic 1: Data Privacy \& Consent in Data Science}

\begin{figure}[h]
\centering
\includegraphics[width=1.0\textwidth]{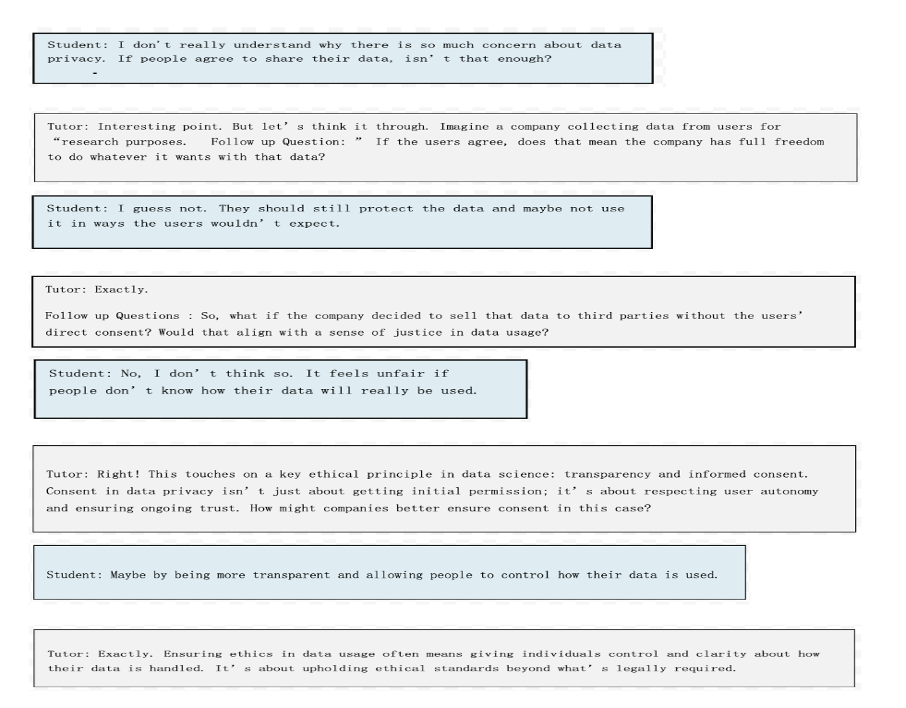}
\caption{Example of Scocratic Dialogue}\label{fig100}
\end{figure}

\subsubsection{Reflection and Feedback Mechanism}\label{subsubsec46}
After each session, students can complete a reflection exercise. 

\textit{Tutors could ask.} "What is one ethical principle you struggled to understand before this session, and how has your understanding changed?" 
The tutor can gather feedback on which Socratic questions led to the most productive discussions, fine-tuning future sessions. 

\subsection{Contextual Output Variation Using Moodle-Embedded Prompts}\label{subsec43}

To evaluate BotCaptain’s ability to personalize responses, we compared AI-generated answers to the same set of questions presented with and without contextual embedding from our iCycle Moodle course. When course-specific materials (e.g., lesson objectives, key vocabulary, or assignment prompts) were embedded within the prompt, the AI produced more targeted and accurate explanations.  

As illustrated in Fig. 11, the embedded prompt response dialogue on the right reflects a more conversational and pedagogically aligned interaction. It engages the learner through questions and scenarios tied to the course context, prompting critical reflection. In contrast, the basic prompt on the left yields a more generic explanation, structured as a list without adapting to the learner's intent or background. This comparison underscores the added instructional value of embedding Moodle course elements into prompts, enabling BotCaptain to deliver adaptive, course-aware feedback that better supports learner understanding. 

\begin{figure}[h]
\centering
\includegraphics[width=0.8\textwidth]{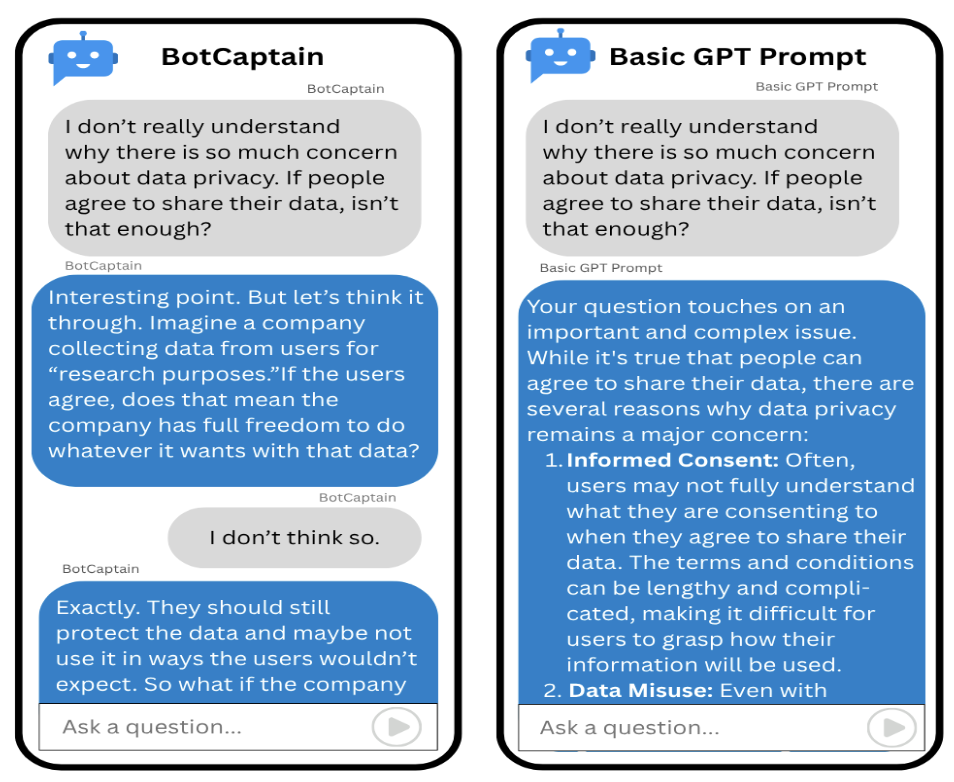}
\caption{Comparison of Embedded vs. Basic AI/GPT Prompt Responses.}\label{fig16}
\end{figure}
The outputs further illustrate how BotCaptain responds to the same student question under two conditions: (A) with embedded contextual prompts tied to the Moodle course and (B) using a basic, non-contextual prompt. In the embedded version, the AI engages the student in a conversational and reflective dialogue, encouraging critical thinking through follow-up questions. The basic version provides a more general explanation in list format without adapting to the learner’s engagement. The comparison highlights the enhanced personalization and pedagogical depth made possible by embedding instructional context in AI prompts. 

Finally, these responses demonstrate improved alignment with the intended learning outcomes and greater relevance to the Data Science course structure. In contrast, prompts lacking embedded context yield more generic responses, underscoring the importance of contextual cues in tailoring AI assistance. 

\subsection{Comparative Discourse Analysis between Learner Interactions}\label{subsec44}
To further assess personalization, we analyzed how BotCaptain adapted its dialogue across different learners responding to the same set of quiz items. Two students with varied levels of understanding and answer patterns engaged with the AI on identical questions. 

From figure 12, the resulting discourse paths diverged significantly, one offering summarization and reinforcement, the other prompting elaborative scaffolding and clarifying questions. This dynamic response behaviour illustrates the AI’s capacity to personalize interactions based on individual learner input, fostering differentiated instruction and more meaningful feedback loops based on learner level of course comprehension. 

\begin{figure}[h]
\centering
\includegraphics[width=0.7\textwidth]{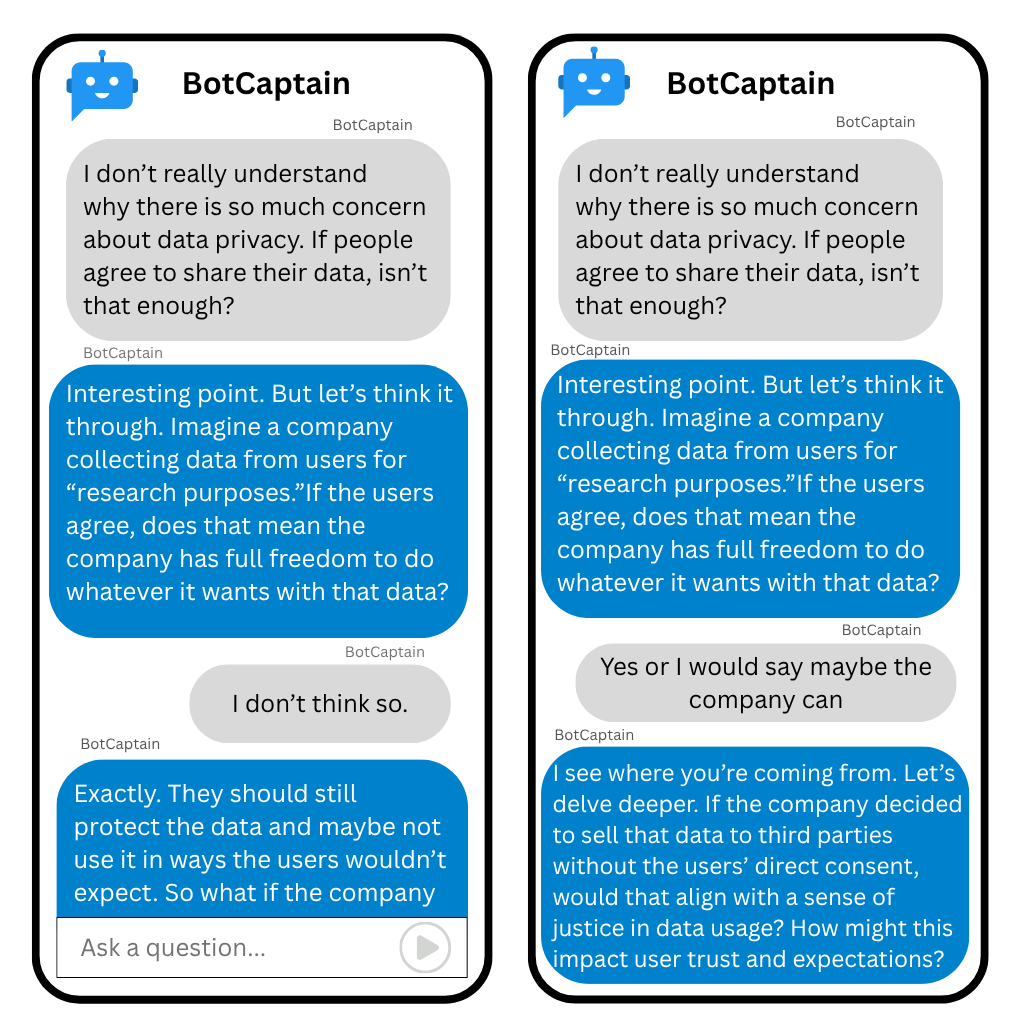}
\caption{Differentiated Dialogue Paths Based on Learner Understanding.\\
The left panel (Learner 1) shows a response pattern with summarization and reinforcement, indicating the learner has a baseline grasp of the topic. The right panel (Learner 2) demonstrates how BotCaptain adapted by prompting deeper inquiry and ethical reflection to support a learner with lower initial understanding. This contrast illustrates the system’s capacity to tailor scaffolding dynamically, making instructional dialogue more responsive to individual learner needs.}\label{fig12}
\end{figure}

\subsection{Pedagogical Impact and Future Enhancements of Chat-Based AI Integration}\label{subsec45}
The deployment of chat-based AI agents such as BotCaptain, integrated within LMS environments, demonstrates transformative potential in enhancing pedagogical effectiveness across multiple dimensions of online education. Moving beyond basic responsiveness such as Chat-GPT, these agents act as dynamic instructional allies, providing formative, adaptive, and reflective learning support. The following pedagogical outcomes are particularly noteworthy.

\subsubsection{Scaffolded Learning Through Conversational Tutoring }\label{subsubsec451}
BotCaptain enables real-time scaffolding by offering contextually relevant prompts and micro-explanations tailored to the learner’s stage of progress. This promotes just-in-time support akin to a human tutor, where the AI dynamically modulates complexity based on interaction history. 

For instance, when students ask vague or under-defined questions, the agent can pose clarifying follow-ups or reformulate the query into a more pedagogically rich version, fostering metacognitive engagement. 

\subsubsection{Personalized Socratic Dialogue and Critical Thinking Stimulation}\label{subsubsec452}
Rather than simply delivering facts, BotCaptain can engage in Socratic questioning, prompting students to reflect more deeply on ethical dilemmas, problem-solving approaches, or conceptual misunderstandings. 

This capability has been especially effective in domains such as data ethics and policy, where AI agents encourage students to weigh trade-offs (e.g., privacy vs. innovation) and articulate reasoned positions as demonstrated in the structured privacy dialogue simulations integrated within our course (see Fig. Topic 1). 

\subsubsection{Seamless Learning Record Integration for Reflection and Analytics}\label{subsubsec453}

Each conversational thread is logged via xAPI-compatible channels, contributing to the learner’s LRS profile. This offers dual benefits: students gain access to a reflective history of inquiries, while instructors can visualize aggregate interaction data to detect common misconceptions, dormant learners, or FAQ hotspots. 

Dashboards can be designed to highlight temporal patterns of confusion and peak engagement periods across modules, enabling data-informed pedagogical decisions. 

\subsubsection{Equity and Accessibility in AI-Supported Learning }\label{subsubsec454}

By offering unbiased, always available, and linguistically adaptive support, the chat agent helps level the playing field for learners with limited access to peer support, tutoring services, or synchronous instruction. 

The interface supports natural language queries without requiring technical jargon, and with future multilingual enhancements, could significantly broaden global access to personalized instruction. 

\section{Benchmarking VITA against Educational AI Hub, GPTutor and Other AI Tutor}\label{sec5}

The benchmark aims to evaluate the capabilities of the VITA-BotCaptain adaptive learning system, particularly as it applies to data science education, by comparing it with both commercial and research-grade intelligent tutoring systems.

\subsection{Purpose of Comparison}\label{subsec51}
Recent systems such as GPTutor (Sajja et al., 2025) have demonstrated the potential of large language models in providing personalized tutoring through retrieval-augmented generation (RAG). 

GPTutor allows instructors to upload course materials and leverages a two-stage embedding pipeline to retrieve relevant content for student queries. Its dashboard features, including leaderboards and follow-up prompts, are designed to enhance engagement and monitor performance. 

The Educational AI Hub (Gul et al., 2023), a comprehensive and open-access deployment designed for STEM education, serves also as the most directly comparable platform due to its end-to-end architecture, modular AI agents, and LTI integration strategy. To assess the effectiveness and technical maturity of the VITA BotCaptain system, we benchmarked it against leading AI-driven educational platforms, including the Educational AI Hub, IBM Watson Tutor, and Duolingo AI, focusing on core capabilities in personalization, feedback quality, system integration, and data interoperability. This benchmark provides a systematic comparison and identifies areas where VITA excels and where future enhancements are needed. 

\subsection{Benchmark Criteria}\label{subsec52}

The comparative evaluation was structured across the following axes described in Table 2 below.

\begin{table}[h!]
\centering
\caption{Benchmark Criteria Definition}
\label{tab:benchmarks}
\begin{tabular}{@{}p{4cm} p{8.5cm}@{}}
\toprule
\textbf{\textit{Dimension}} & \textbf{\textit{Description}} \\
\midrule

\textbf{Adaptive Pathways} & \parbox[t]{8.5cm}{Ability to customize learning flow based on \\
learner profile/performance} \\ \\

\textbf{LMS Integration} & \parbox[t]{8.5cm}{Native compatibility with LMSs (e.g., Moodle,\\
Canvas) via LTI or plugin} \\ \\

\textbf{xAPI and LRS Compatibility} & \parbox[t]{8.5cm}{Ability to generate and transmit learning\\
records via xAPI to an LRS} \\ \\

\textbf{Hallucination Mitigation} & \parbox[t]{8.5cm}{Mechanisms to detect or reduce inaccurate \\
LLM outputs.} \\ \\

\textbf{Retrieval-Augmented Generation} & \parbox[t]{8.5cm}{Use of embedded knowledge bases or \\
RAG pipelines for grounded reasoning.} \\\\

\textbf{Math \& Code Intelligence} & \parbox[t]{8.5cm}{Capacity to handle symbolic content, e.g.,\\
LaTeX, Python, or math equations} \\ \\

\bottomrule
\end{tabular}
\end{table}

\subsection{Evaluation Framework for VITA \& GPTutor}\label{subsec53}
The evaluation framework assesses four broad domains to evaluate the effectiveness and robustness of the systems. 

\subsubsection{Pedagogical Objective}\label{subsubsec531}
The first domain, personalization and pedagogy, focuses on the use of adaptive learning pathways, the quality of feedback provided to learners, and the system’s capacity for fostering emotional engagement.
\begin{table}[h]
\caption{Pedagogical Objective Comparison Table}\label{tab1}%
\begin{tabular}{@{}llll@{}}
\includegraphics[width=0.9\textwidth]{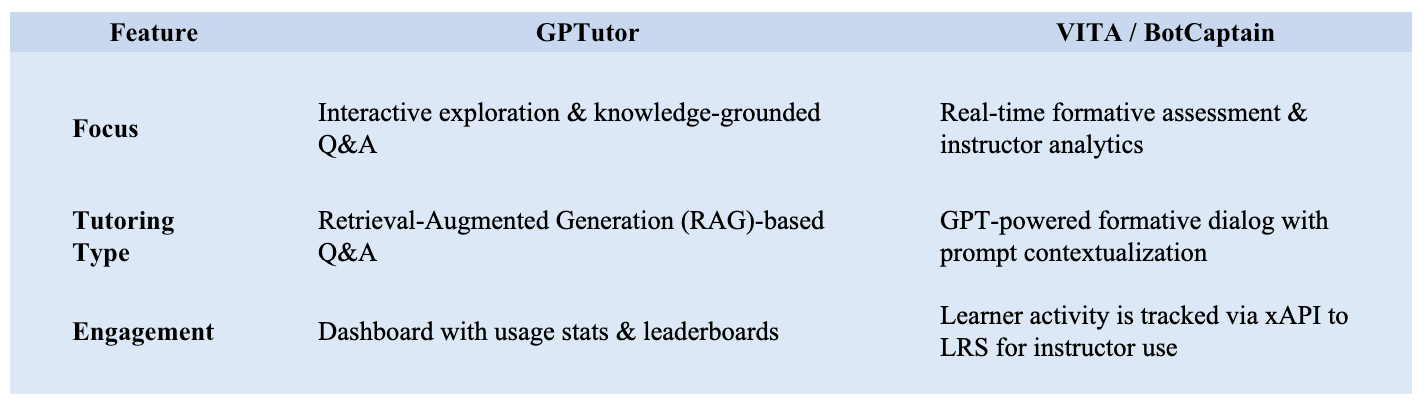}
\end{tabular}
\end{table}

\subsubsection{Architecture \& Customization}\label{subsubsec532}

The second domain, architecture and interoperability in Table 4., examines how well the system integrates with learning management systems (LMSs), supports standards such as xAPI and LTI, and maintains modularity for scalability and customization.

\begin{table}[h]
\caption{Architecture \& Customization Comparison Table}\label{tab2}%
\begin{tabular}{@{}llll@{}}
\includegraphics[width=0.9\textwidth]{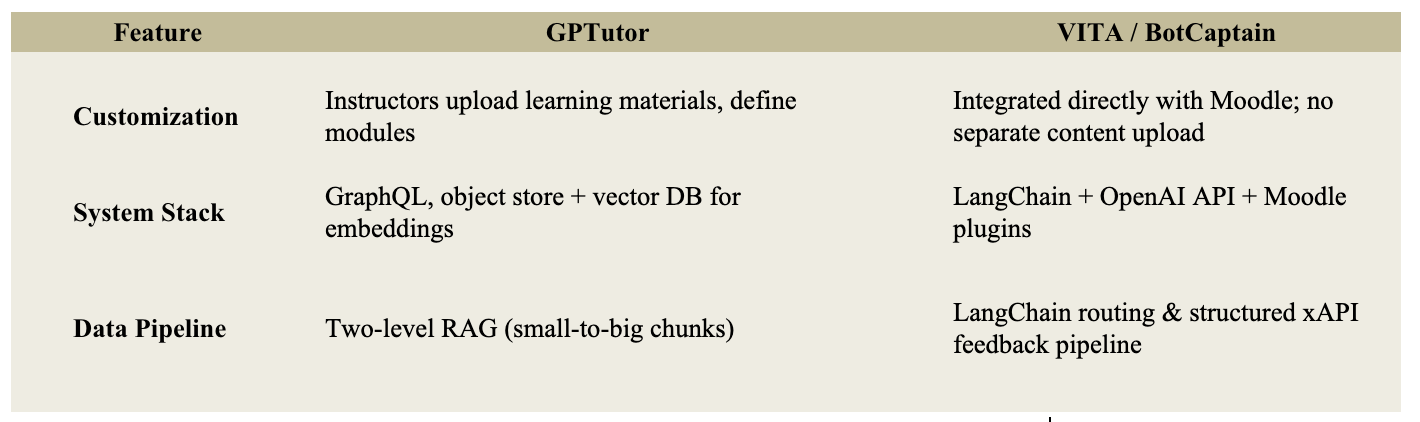}
\end{tabular}
\end{table}

\subsubsection{Evaluation \& Impact}\label{subsubsec533}

The third domain, AI engine and trustworthiness in Table 5., addresses the reliability of the AI through prompt engineering techniques, mechanisms for filtering hallucinated content, and validation of generated responses.

\begin{table}[h]
\caption{Evaluation \& Impact Comparison Table}\label{tab5}%
\begin{tabular}{@{}llll@{}}
\includegraphics[width=0.9\textwidth]{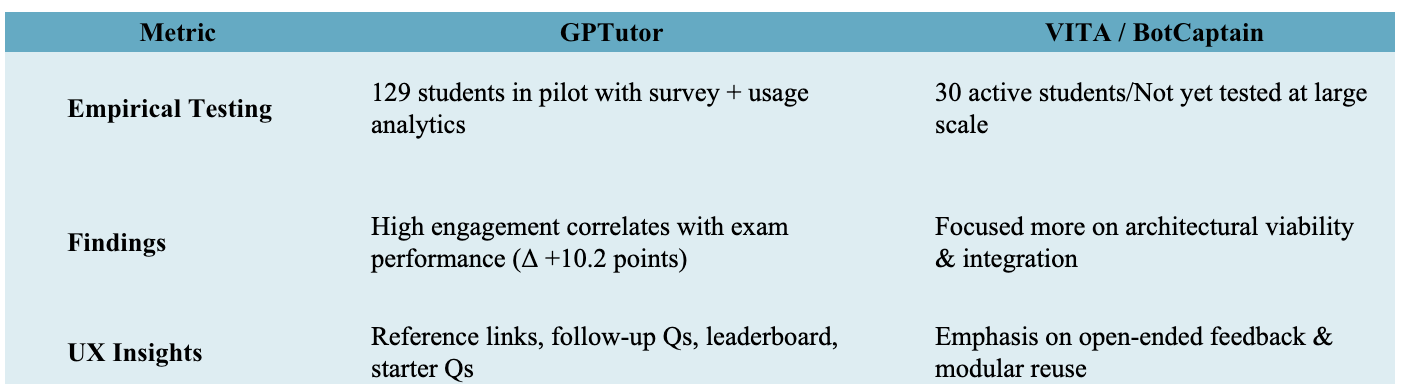}
\end{tabular}
\end{table}

\subsubsection{Articial Intelligence Strategy }\label{subsubsec534}

Lastly, the STEM Capability domain in Table 6. evaluates the system’s support for mathematical problem-solving, programming, and symbolic reasoning, ensuring it can effectively assist in science, technology, engineering, and mathematics education.

\begin{table}[h]
\caption{AI Strategy Comparison Table}\label{tab3}%
\begin{tabular}{@{}llll@{}}
\includegraphics[width=0.9\textwidth]{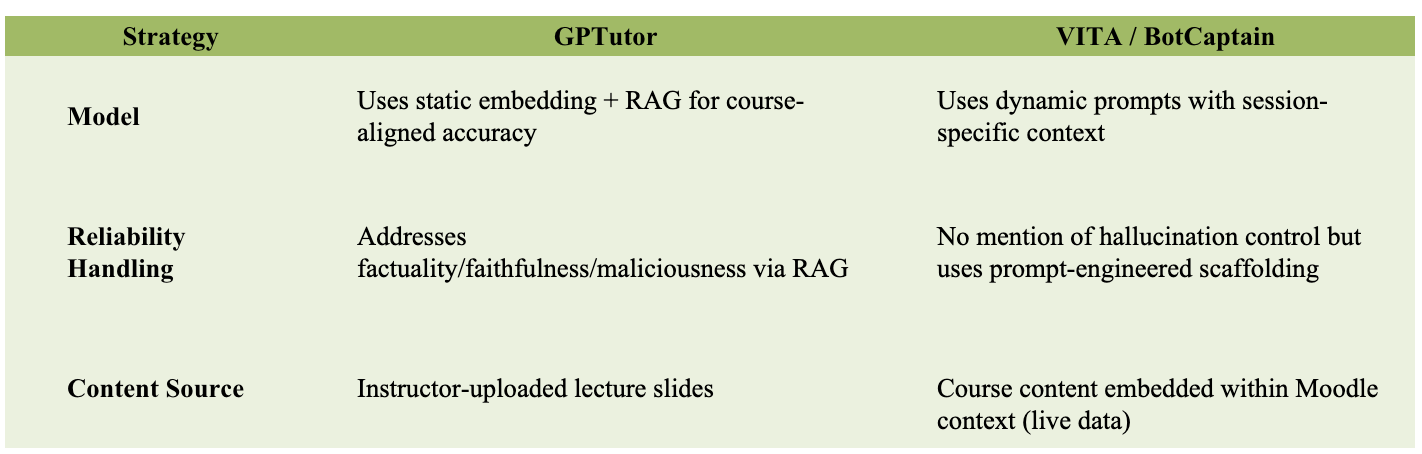}
\end{tabular}
\end{table}

\subsubsection{Strategic Difference Summary}\label{subsubsec535}
GPTutor prioritizes content fidelity and retrieval accuracy, aiming to anchor student queries to uploaded course content using modern RAG techniques. 

VITA/BotCaptain aims to offer real-time, dialog-based AI tutoring tightly coupled with LMS systems, emphasizing open-source extensibility and pedagogical analytics via xAPI. 

In contrast, our VITA-BotCaptain system focuses on formative assessment within a live LMS environment, offering immediate contextual feedback without requiring pre-uploaded content. VITA integrates directly with Moodle and Veracity LRS, enabling xAPI-based tracking of learner interactions and instructional analytics. While GPTutor prioritizes retrieval accuracy and anchored responses, VITA emphasizes modularity, open-source adaptability, and instructor-informed feedback loops.  
 
Together, these two systems represent complementary directions: GPTutor as a content-grounded tutoring assistant, and VITA as a real-time pedagogical co-pilot embedded within institutional infrastructure. 

\subsection{VITA, AI Hub, Duolingo and IBM Tutor Comparison Table}\label{subsec54}

\begin{table}[h]
\caption{Comparative Evaluation Table of VITA, AI Hub, Duolingo and IBM Tutor}\label{tab4}%
\begin{tabular}{@{}llll@{}}
\includegraphics[width=0.9\textwidth]{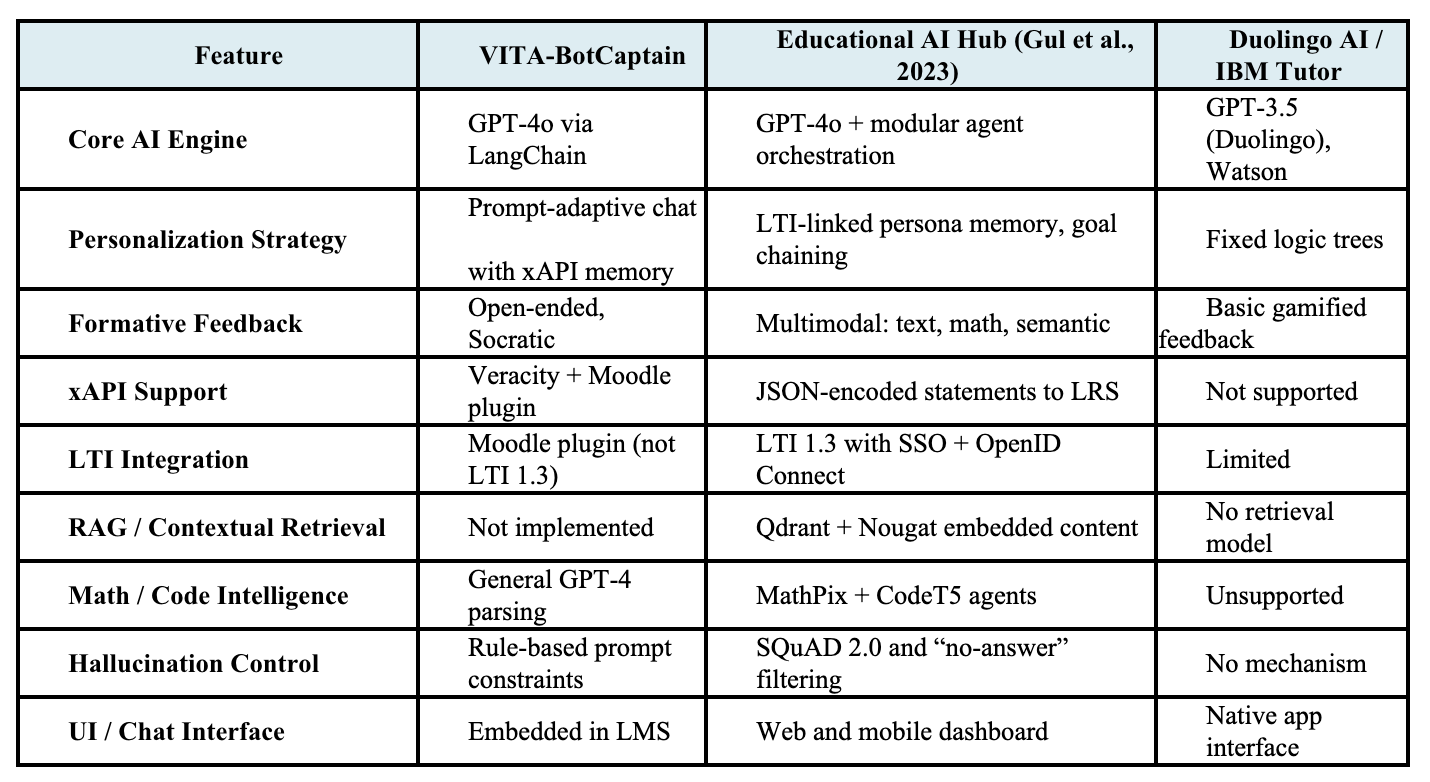}
\end{tabular}
\end{table}

\subsubsection{Analysis of VITA and Educational AI Hub}\label{subsubsec541}
While the Educational AI Hub and VITA both leverage OpenAI’s GPT technology, their deployment philosophies differ: 
\begin{itemize}
    \item VITA emphasizes modularity, ease of integration with Moodle, and rapid deployment of formative prompt-based learning. 
    \item The AI Hub, however, is more mature in security (LTI 1.3, OIDC), retrieval-based answering, and response trustworthiness, leveraging Qdrant vector stores, MathPix OCR, and agent-based memory routing. 
\end{itemize}

Our VITA-BotCaptain platform demonstrates competitive strengths in formative feedback delivery, LMS integration, and LRS connectivity. Its lightweight architecture allows easy deployment in open-source LMS environments, particularly Moodle. However, the platform currently lacks a formal RAG engine for content grounding and does not yet implement robust hallucination filtering, both of which are addressed in the Educational AI Hub via Nougat parsing and no-answer QA evaluation using GPT-4o. 

\begin{figure}[h]
\centering
\includegraphics[width=0.8\textwidth]{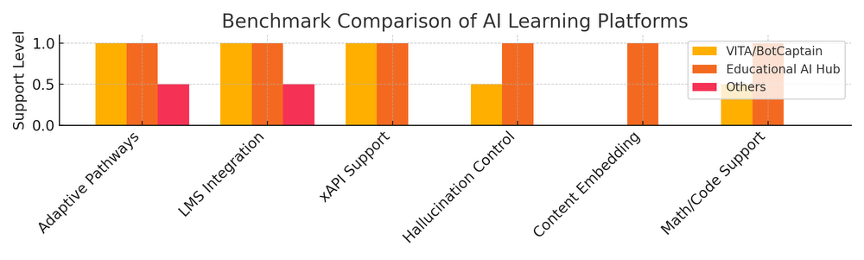}
\caption{Graphic Benchmark Comparison of AI Learning Plarforms}\label{fig17}
\end{figure}

Notably, the Educational AI Hub leverages LTI 1.3 login protocols, SSO tokens, and OpenID Connect, offering seamless LMS integration and fine-grained analytics access. In contrast, BotCaptain uses a plugin-based approach with xAPI statements stored in Veracity LRS, which while modular may limit portability across other learning ecosystems without LTI certification. 

Furthermore, the AI Hub supports multi-agent coordination, with components like MathAgent, CodeAgent, and PersonalAgent communicating over a unified graph memory features that could significantly enhance VITA if adopted.

\subsubsection{Strategic Positioning of VITA}\label{subsubsec542}
While commercial platforms prioritize polish and UX, and the AI Hub leads in research-grade architecture, VITA occupies a crucial middle space: open-source, educator-controlled, and rapidly customizable. Its integration with Moodle and Veracity LRS makes it uniquely suited for institutions seeking control, affordability, and extensibility. With the above enhancements, VITA could evolve into a full-spectrum AI tutor-as-a-service framework for STEM and data science education. 

Moreover, unlike proprietary solutions that often operate as closed systems with limited adaptability, VITA offers a transparent and modular ecosystem that empowers educators to design, monitor, and iterate on AI-assisted learning workflows. Its plug-and-play compatibility with open standards such as xAPI and LTI ensures interoperability with diverse LMS environments, while its lightweight deployment footprint enables rapid prototyping in institutional contexts without vendor lock-in. VITA’s architecture not only supports course-level personalization but also institutional analytics, allowing for scalable deployment across departments or curriculum. With proper investment in multilingual support, bias mitigation, and domain-specific fine-tuning, VITA could become the reference implementation for AI-powered instructional agents in open education, bridging the gap between cutting-edge research and real-world educational equity.

\subsection{Proposed Recommendations and Integration Pathways for VITA}\label{subsec55}

To further enrich the learning experience, the next phase of development envisions: 
\begin{itemize}
    \item \textbf{RAG-based} (Retrieval-Augmented Generation) Chat Agents, drawing from curated course documents or instructor-defined ground truths to mitigate hallucination risks. 
    \item \textbf{Confidence-Level Tagging}, where the AI transparently indicates the certainty of its answers, empowering students to cross-validate and cite appropriately. 
    \item \textbf{Instructor Feedback Loops}, enabling teachers to edit or annotate AI responses in situ, creating a co-teaching paradigm. 
    \item \textbf{Integration with Assessment Platforms}, where the agent not only answers but helps students unpack rubric-aligned expectations, scaffolds assignment planning, or runs mock oral defense simulations.
\end{itemize}

To close the gap and position VITA-BotCaptain as a cutting-edge educational assistant, we propose the following enhancements: 
\begin{itemize}
    \item \textit{Upgrade to LTI 1.3 + OIDC:} For secure and portable LMS integration across platforms like Canvas and Blackboard.
    \item \textit{Implement Formal Evaluation Metrics:} Use SQuAD 2.0 and BLEU-style metrics to monitor LLM hallucination, especially for knowledge-based or technical queries. 
    \item \textit{Embed OCR Agents:} Add lightweight agents (e.g., MathPix or open-source Nougat) to parse equations, charts, or scanned PDFs. 

    \item \textit{Implement No-Answer Detection:} Incorporate response confidence thresholds and filters based on SQuAD 2.0 to detect hallucinations. 
    \item \textit{Introduce Memory Agents:} Modular personas or skill-specific agents to personalize feedback based on learner context.
    \item \textit{Incorporate Math-Aware Parsing:} Integrate MathPix API or open-source LaTeX parsers to handle mathematical content in formative tasks. 
\end{itemize}
                                                    
These improvements would elevate the VITA platform’s fidelity, extend its interoperability, and bring it into closer alignment with leading research-based deployments like the Educational AI Hub.

\section{AI-Enhanced Learning Assessment and Actionable Insights}\label{sec6}
This section presents visual learning analytics dashboards and insights, showcasing the clarity of online learning activities accessible via the dashboard (https://erau.xapi.io). To ensure the integrity of online education, we designed formative learning assessments that mitigate the extent to which students can misuse ChatGPT in an effective and reliable manner.

\subsection{Mitigate misusing ChatGPT or OpenAI for quizzes or Homework Assignments}\label{subsec61}

We developed four approaches that can be incorporated into the design of assessments for DS courses. These approaches leverage the fact that ChatGPT is not designed to answer queries based on information outside of its training dataset.  We present examples of the four approaches and address practical issues that might limit the implementation of these approaches in current online assessment platforms. 

The first class of assessments, “self-referential queries”, are conditioned on prior information covered in the class of which ChatGPT does not have direct knowledge.  Example 1– A study is performed to test if rats are empathetic. The study involves recording whether one rat helps another. The class learns how to formulate a vague question, “are rats empathetic”, into a statistical question about the value of a parameter. The assignment then asks students to apply this same approach to data from a poll on whether a person feels their significant other’s use of social media affects their relationship, using the same approach used in the rat empathy study. 

The second class of assessments, “information-starved queries”, leverage idiosyncratic approaches an educator uses, which are not part of the mainstream curriculum., “information-starved queries,” leverage idiosyncratic approaches an educator uses which are not part of the mainstream curriculum.  Example 2: You may define a statistical question as a question about the value of a parameter that requires 4 pieces of information to be “well defined” (purpose, population, variable, and parameter). The assignment asks students to identify the four parts of a statistical question and then formulate it.  

The third class of assessments, “temporally based queries”, uses the fact that ChatGPT is not trained on current data, though ChatGPT 4.0 has improved its time-sensitive information. These assignments ask students to analyze a current dataset such as a local sports report from the current week or current day’s stock market report.  

The fourth approach is to assign students the Socratic dialogue shown in section 4.2 and use the embedded VITA to evaluate the students’ performance. Example 3: Instead of asking students to answer a DS ethic problem, we ask the students to start a Socratic dialogue about the misconducts and consequences. We can ask another student to analyse the logical correctness and flaws of the dialog.  

\subsection{Overview to the Learning Activities}\label{subsec62}

The LRS URL for Data Visualization is (https://erau.xapi.io/default-lrs/xapi/). A CSV file named as “statementViewerExport” can be downloaded for all xAPI records (see the Appendix for the link to access the file). The file has four columns, representing the Actor (who), Verb (did), Object (what), and the activity Timestamp (when), and 206264 records, representing the number of xAPI statements collected so far in reverse chronical order. The LRS URL for is https://erau.xapi.io/data-mining/xapi/ that collected 140,723 xAPI statements.  

The predictive models for learning analytics based on the learning activity datasets in LRS can assist instructors and TA's in catching early indicators of either success or failure in a course.  

The visualization below is a crucial component of evaluating the learning process at scale. Most LMS and Auto Tutor tools such as GIFT allow the instructor to check the progress of individual learners and a summative data of a whole class. This is time-consuming if the instructor needs to check each learner's progress periodically. This process of notifying the instructor is automated by the machine learning models running in the background. Instead of checking the records of each student, the instructor can configure the system to send only the results of underperforming outliers for the instructors to intervene as deemed necessary.  

Figure 11 illustrates the collective learning activities, sorted from the most active to the least active learners. This graph helps instructors identify outliers at both the positive and negative ends of the spectrum.  
\subsection{Visualizing the outliers for underperformed students}\label{subsec63}

The instructor can select under-performing students to view the dashboard in Figure 11, which provides an overview of an individual learner’s activities and content details. This allows instructors to pinpoint specific areas where the student is falling behind and offer constructive feedback and support.

\begin{figure}[h]
\centering
\includegraphics[width=0.8\textwidth]{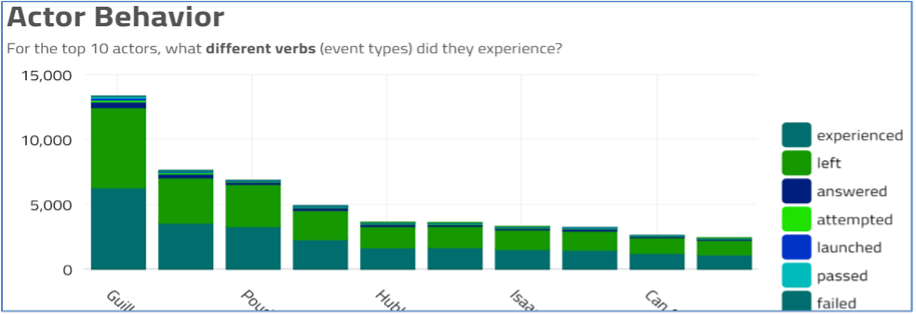}
\caption{Overview of the collective learning activities sorted by learners}\label{fig11}
\end{figure}

\begin{figure}[h]
\centering
\includegraphics[width=0.8\textwidth]{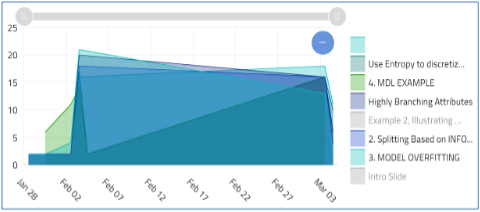}
\caption{Overview of an individual learner’s learning activities and content}\label{fig122}
\end{figure}

Figures 13 and 14 illustrate the number of meetings attended, the number of quizzes passed per week by each learner, and the summative counts of Passed or Failed for each quiz broken down by learners and weeks.  Based on the data shown in Figure 6, the dashboard allows for interaction to filter both actors and date ranges and help the instructor gain insight into which students are performing well and which need intervention. Such observations help the instructor grade team projects fairly, rather than equally distributing grades, which can undermine motivation and encourage hitchhiking and social loafing. Consequently, the formative assessment data motivate learners to actively take on leadership roles and make significant contributions to the project's success. 

\begin{figure}[h]
\centering
\includegraphics[width=0.8\textwidth]{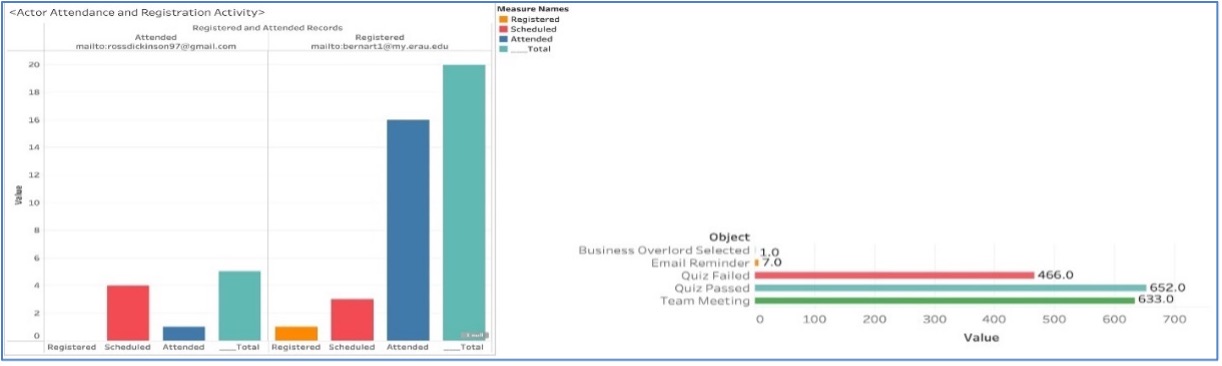}
\caption{Number of meetings attended and registrations by each learner }\label{fig13}
\end{figure}

\begin{figure}[h]
\centering
\includegraphics[width=0.8\textwidth]{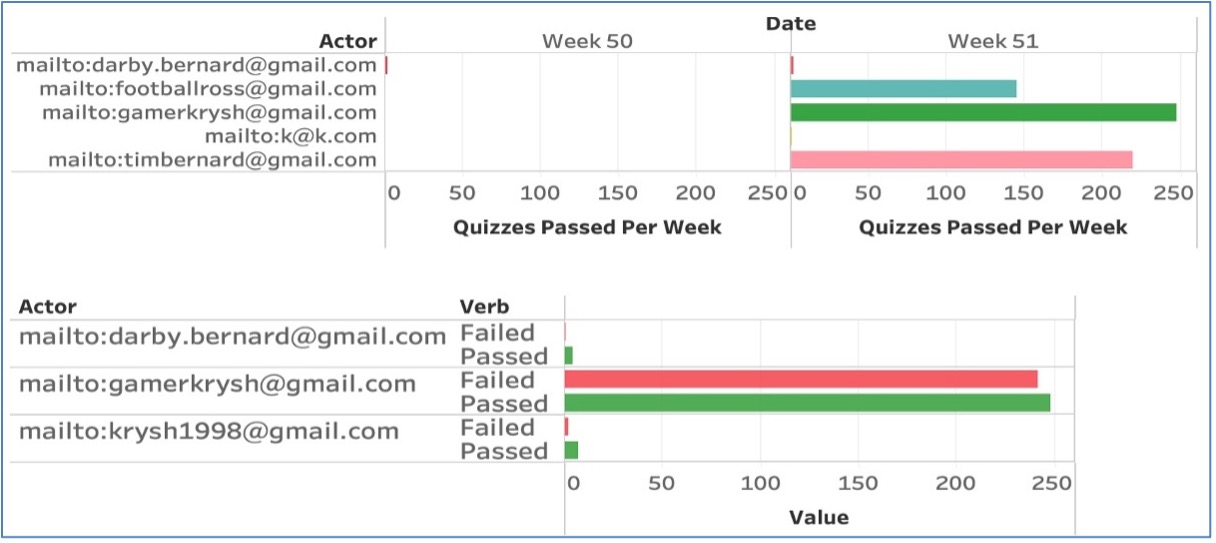}
\caption{Counts of Passed or Failed for each quiz broken down by learners and weeks}\label{fig14}
\end{figure}

\section{Adaptive Learning Pathways for Continuous Personalization}\label{sec7}
This section showcases how formative assessments drive content recommendations, guiding learners through progression, reinforcement, or remediation paths based on their competence. Learner competence is evaluated using formative assessments and course prerequisites to recommend new content. As shown in Figure 15, there are three possible recommendations based on this evaluation: advancing to new content, reinforcing previous content, or revisiting remediation material from earlier courses.

\subsection{Description of the Adaptive Learning Implementation}\label{subsec71}

The Adaptive learning systems we are developing with GIFT, Articulate 360, and Moodle leverage formative assessments to tailor educational content to individual learner needs. Below are detailed examples and expanded explanations based on the adaptive learning framework: 

\subsubsection{Adaptive Learning Engine Configuration}\label{subsubsec711}

The Adaptive Learning Engine Configuration governs how personalized learning pathways are dynamically assigned based on real-time learner performance. This system combines formative assessment data with branching logic to ensure that each student receives content tailored to their level of understanding and progress. 

\begin{figure}[h]
\centering
\includegraphics[width=0.8\textwidth]{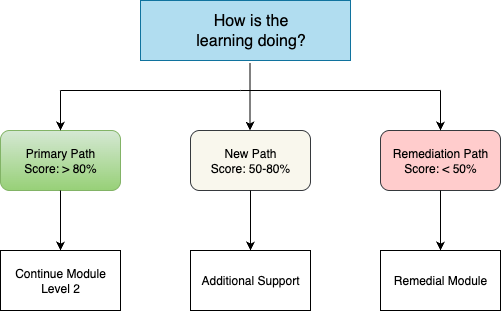}
\caption{Adaptive Learning Pathway Based on Learner Performance}\label{fig15}
\end{figure}

\subsubsection{Formative Assessments \& Dynamic Content Recommendations}\label{subsubsec712}

Regular checkpoints are integrated within course content to assess learner understanding and skills. Based on assessment results, the system automatically suggests one of three paths: 

\begin{itemize}
    \item \textbf{Progression/Primary Path:} Learners who meet or exceed the competency thresholds are directed to advanced modules to continue building on their knowledge. 
    \item \textbf{Reinforcement/New Path:} Learners who show adequate understanding but could benefit from further practice are recommended additional exercises related to current learning topics.
    \item \textbf{Remediation:} Learners who do not meet the required thresholds are guided to review modules that address foundational knowledge gaps.
\end{itemize}

\subsubsection{Branching Logic Details}\label{subsubsec713}

Branching Logic Details outline the underlying mechanisms that enable the adaptive engine to tailor learning experiences based on real-time learner inputs and instructor oversight. 
\begin{itemize}
    \item \textbf{Data-Driven Path Selection:} Utilizes an algorithmic approach where learner responses are analyzed in real-time to determine the most suitable learning path.
    \item \textbf{Instructor Oversight and Override:} While the system automates path selection, instructors can manually adjust paths based on their assessment of learner performance and engagement. 
    \item \textbf{Content Preparation:} Development of a rich database of content varying in difficulty and depth, allowing for seamless transition between different paths as dictated by learner performance. 
\end{itemize}

\subsection{Overview of the System Implementation}\label{subsec72}

This section outlines the core components of the system’s implementation, emphasizing how adaptive logic and learner data are leveraged to deliver a personalized educational experience. Drawing on scenario-based interactions, real-time feedback, and long-term performance tracking, the system intelligently adjusts content delivery to align with individual learner needs, progress, and professional goals. 

\begin{itemize}
    \item \textit{Scenario-Based Learning:}
    \begin{itemize}
    \item Contextual Branching: Learners are placed in simulated scenarios relevant to the course material. Decisions made within these scenarios determine the subsequent content, emphasizing practical application of knowledge. 
    \item Feedback Loops: Immediate feedback is provided for choices made during scenarios, with explanations and recommendations for further study. 
\end{itemize}
\end{itemize}

\begin{itemize}
    \item \textit{Skill-Based Branching: }
    \begin{itemize}
    \item Skills Inventory: At the start of the course, learners complete a skills inventory which helps to map out a personalized learning path.  
    \item Adaptive Quizzes: Frequent quizzes adapt in complexity based on the learner’s past performance, ensuring that they are continually challenged but not overwhelmed. 
\end{itemize}
\end{itemize}

\begin{itemize}
    \item \textit{Long-Term Learning Path Optimization:}
    \begin{itemize}
    \item Performance Tracking: This feature is a work in progress. Over the course of several terms, learner performance data is accumulated and analyzed to refine the adaptivity algorithms, improving the accuracy of path assignments based on the data saved in the transactional LRS.
    \item Integration with Career Goals: For learners in vocational or professional tracks, the learning path suggestions can also consider the skills and knowledge most relevant to their career aspirations based on the data saved in the authoritative LRS. This feature will be part of our future work.
\end{itemize}
\end{itemize}

The branching logic as shown in Fig. 19, though straightforward technically, relies on the instructor's expertise and is time-consuming due to content preparation. We implemented this feature in the GIFT system as a proof of concept and plan to invest the next two years in refining and integrating this feature in both GIFT and Moodle. 

\begin{figure}[h]
\centering
\includegraphics[width=1.0\textwidth]{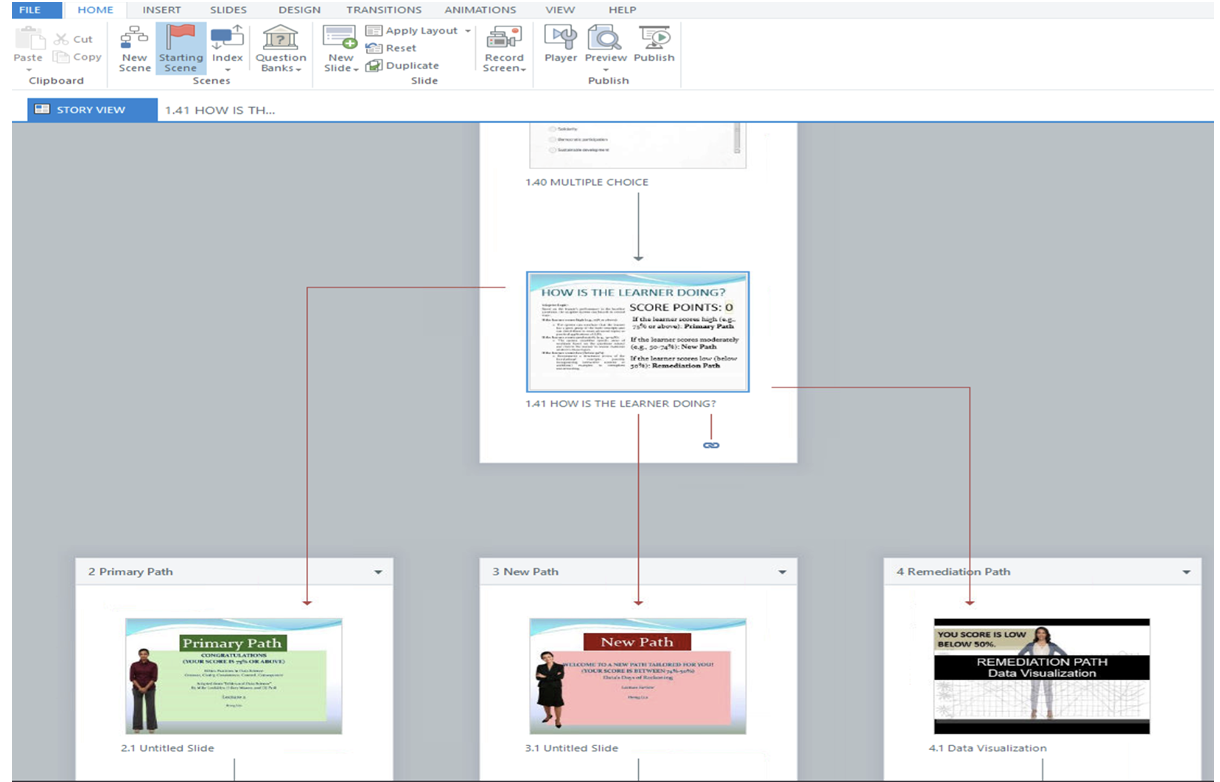}
\caption{Adaptive Content Recommendation based on Learning Assessment (Articulate 360)}\label{fig19}
\end{figure}

\section{Discussion}\label{subsec8}

Our three-pathway system (progression, reinforcement, remediation) shows that effective AI personalization works best when algorithms make initial recommendations, but instructors retain final decision-making authority. The system can automatically suggest learning paths based on assessment data, which helps address the scalability problems in data science education. But the instructor override feature proved crucial there are simply too many individual circumstances that algorithms cannot account for. 

This implementation tackles several gaps we identified in existing research. Ferguson (2012) noted the persistent disconnect between what educational technology can do and how it is used in teaching. Our approach bridges this gap by analyzing not just whether students get answers right, but how their thinking develops through dialogue with the AI. This provides instructors with much richer information about student learning than traditional quiz scores or clickstream data. 

Our strategies for preventing AI misuse revealed something interesting about changing educational priorities. As AI becomes better at routine information processing, the value of education increasingly lies in developing capabilities that AI cannot replicate critical thinking, ethical reasoning, and especially the ability to ask good questions. Our Socratic dialogue feature worked well because it guided students toward discovery rather than just providing answers. Students seemed to develop stronger analytical skills when the AI challenged them to think rather than simply giving them information. 

\section{Conclusion and Future Directions}\label{sec9}

This paper highlights the transformative potential of integrating OpenAI’s virtual teaching assistant, BotCaptain, with adaptive learning technologies to revolutionize personalized education. The automated formative assessments, adaptive learning pathways, and visual analytics dashboards demonstrate how conversational AI can elevate modern learning management systems. These innovations not only streamline instructional processes but also deliver tailored learning experiences that foster engagement, critical thinking, and deeper comprehension. 

The ADL platform, iCycle, and its virtual TA are at an exciting stage of early development, building on over a decade of related research efforts. Currently, its use is focused on academic research, with plans to scale for broader commercial applications. A forthcoming article will explore a blockchain-based business model designed to sustain iCycle and its courseware beyond the anticipated conclusion of grant funding in the next one to two years. In the interim, our team is committed to leveraging remaining resources to enhance user experience and expand platform capabilities. 

The integration of BotCaptain, refined using OpenAI’s LLM, into Moodle has already demonstrated its efficacy in delivering real-time contextual support, dynamic formative assessments, and actionable predictive insights. The platform’s ability to identify learning gaps and recommend personalized content positions it as a versatile tool adaptable to diverse educational needs. Additionally, the incorporation of Socratic dialogue and AI-driven prompts promotes critical thinking and active participation, enriching the learning process. 

While the system shows significant promise, challenges remain, particularly in scaling adaptive content preparation and refining Socratic dialogue mechanisms. Future efforts will prioritize addressing these challenges, expanding platform integration across more courses, optimizing predictive analytics, and leveraging authoritative LRS records to recommend new courses.  

In summary, this work underscores the potential of AI-driven solutions to transform education by making it more flexible, scalable, and learner-centered. By addressing current challenges and building on its successes, the ADL platform is poised to play a pivotal role in shaping the future of personalized learning.  Our future work will focus on scaling adaptive content preparation, refining Socratic dialogue, and making the platform available to other teachers and students from other science, technologies, engineering, and mathematics disciplines.  




\backmatter

\bmhead{Acknowledgements}

 The authors thank Veracity Learning for donating the Learning Record Store (LRS) software and the GIFT team for technical support. This work was supported by the U.S. National Science Foundation (NSF). Any opinions,
findings, and conclusions or recommendations expressed in this material are those of the authors and do not necessarily reflect the views of the NSF.



\section*{Declarations}


\begin{itemize}
\item Funding:
This research was supported by the U.S. National Science Foundation (NSF) under Grant Nos. 2142514, 2142327, and 2142465.
\item Author contribution:
All authors contributed to the study conception and design. Material preparation, data collection and analysis were performed by Fadjimata I. Anaroua, and Hong Liu. The first draft of the manuscript was written by Fadjimata I. Anaroua and Hong Liu and all authors commented on previous versions of the manuscript. All authors read and approved the final manuscript.”
\end{itemize}

\section*{References List}

\hspace*{1.3em} 1. Advanced Distributed Learning. (2018). DoD’s Advanced Distributed Learning Initiative: Total Learning Architecture: 2018 Reference Implementation Specifications and Standards (W900KK-17-D-0004). https://adlnet.gov/projects/tla.
 
2. Anaroua, F. I., Li, Q., \& Liu, H. (2024). Enhancing data science courses pedagogy through GIFT-enabled adaptive learning pathways. Proceedings of the 12th Annual GIFT Users Symposium, Orlando, May 2024.

3. Bienkowski, M., Feng, M., \& Means, M. (2012). Enhancing teaching and learning through educational data mining and learning analytics: An issue brief. Center for Technology in Learning, SRI International, U.S Education Department. https://files.eric.ed.gov/fulltext/ED611199.pdf 

Bloom, B. S. (1984). The 2 sigma problem: The search for methods of group instruction as effective as one-to-one tutoring. Educational researcher, 13(6), 4-16. 

5. Brenna, F., Danesi, G., Finch, G., Goehring, B., \& Goyal, M. (2018). Shifting toward enterprise-grade AI: Resolving data and skills gaps to realize value. IBM Institute for Business Value. https://www.ibm.com/downloads/cas/QQ5KZLEL 

6. Matthew Brenneman, Hong Liu, Assessment Design for Mitigating Academic Misconduct with ChatGPT in Statistics and Data Science Courses, The Twentieth International Conference on Technology, Knowledge, and Society, Theme, People, Education, and Technology for a Sustainable Future, March 7 – 8, 2024, Valencia, Spain. 

7. Bretag, T., Harper, R., Rundle, K., Newton, P. M., Ellis, C., Saddiqui, S., \& van Haeringen, K. (2020). Contract cheating in Australian higher education: a comparison of non-university higher education providers and universities. Assessment \& Evaluation in Higher Education. 

8. Chen, X., Zou, D., Xie, H., \& Cheng, G. (2021). Twenty years of personalized language learning. Educational Technology \& Society, 24(1), 205-222. 

9. Corbett, A. T., Anderson, J. R., \& O’Brien, A. T. (2012). Student modeling in the ACT programming tutor. In Cognitively diagnostic assessment (pp. 19-41). Routledge. 

10. Ferguson, R. (2012). Learning analytics: drivers, developments and challenges. International journal of technology enhanced learning, 4(5-6), 304-317. 

11. FitzGerald, E., Kucirkova, N., Jones, A., Cross, S., Ferguson, R., Herodotou, C., ... \& Scanlon, E. (2018). Dimensions of personalisation in technology‐enhanced learning: A framework and implications for design. British Journal of Educational Technology, 49(1), 165-181. 

12. Burmester, E. (2020). Authoring collective training demonstrations in GIFT. Proceedings of the 8th Annual GIFT Users Symposium (GIFTSym8). 

13.  Doughty, P. L., Spector, J. M., \& Yonai, B. (2003). Time, efficacy, and cost considerations of e-collaboration in online university courses. Brazilian Review of Open and Distance Learning, 2(1). http://www.abed.org.br/publique/cgi/cgilua.exe/sys/start.htm 

14. Edison. (2017). Edison Data Science Framework: Part 1. Data Science Competence Framework (CF-DS). https://creativecommons.org/licenses/by/4.0/ 

15. Gašević, D., Dawson, S., \& Siemens, G. (2014). Let's not forget learning analytics are about learning. https://www.researchgate.net/publication/269999021 

16. Gagné, R. M. (1985). The conditions of learning and theory of instruction (4th ed.). CBS College Publishing. 

Gul, A., Majeed, A., Rafique, M., Rana, H. A., Ullah, A., \& Malik, A. W. (2023). End-to-End Deployment of the Educational AI Hub for Personalized Learning and Engagement. [arXiv preprint arXiv:2306.17113]. 

17. IBM Analytics, (2020). The data science skills competency model: A blueprint for the growing data scientist profession. https://www.ibm.com/downloads/cas/7109RLQM

18. Imogen Casebourne and Rupert Wegerif, December 2024, Book Chapter, The Role of AI Language Assistants in Dialogic Education for Collective Intelligence, “Artificial Intelligence in Education: The Intersection of Technology and Pedagogy,” Editors:  Peter Ilic, Imogen Casebourne, and Rupert Wegerifpage, 111-125.  

19. Lang, C., Siemens, G., Wise, A., \& Gasevic, D. (Eds.). (2017). Handbook of learning analytics. 

20. Laurillard, D. (2013). Rethinking university teaching: A conversational framework for the effective use of learning technologies. Routledge. 

21. Liu, H., Ludu, A., Klein, J., Spector, J. M., \& Ikle, M. (2017). Innovative model, tools, and learning environments to promote active learning for undergraduates in computational science \& engineering. Journal of Computational Science Education, 8(3), 11-18. https://doi.org/10.22369/issn.2153-4136/8/3/2 

22. Liu, H., Spector, J. M., \& Ikle, M. (2018). Computer technologies for model-based collaborative learning: A research-based approach with initial findings. Computer Applications in Engineering Education, 1-10. https://doi.org/10.1002/cae.22049 

23. Liu, H., Warner, T., Ikle, M., \& Mitten, S. (2020). Harness big data by iCycle - Intelligent computer-supported hybrid collaborative learning environment. International Journal of Smart Technology and Learning, 2(1), 31-47. 

24. Liu, H., Bernard, T., \& Acharya, K. (2020). Using GIFT to develop adaptive remedial courses for graduate degree programs in data science. Proceedings of the 8th Annual GIFT Users Symposium, 61-68. 

25. Hong Liu, Tim Bernard, Elif Cankaya, Alex Hall, (2023), Task-Agnostic Team Competence Assessment and Metacognitive Feedback for Transparent Project-Based Learning in Data Science, to appear on the Int. Journal of Smart Learning Technology, 137-162, retrievable at https://www.inderscienceonline.com/doi/pdf/10.1504/IJSMARTTL.2023.129623. 

26. Liu, H., Malone, N., Yedjou, C., Chadwick, R., Haag, J., \& Spector, M. A. (2024). Automating formative assessment for STEM courses in hybrid learning environments. 2024 IEEE Global Engineering Education Conference (EDUCON), Kos, Greece, 2024. 

Lui, R. W. C., Bai, H., Zhang, A. W. Y., \& Chu, E. T. H. (2024). GPTutor: A Generative AI-powered Intelligent Tutoring System to Support Interactive Learning with Knowledge-Grounded Question Answering. 2024 International Conference on Advances in Electrical Engineering and Computer Applications (AEECA), 702–707. https://doi.org/10.1109/AEECA62331.2024.00124 

27. Merrill, M. D. (2013). First principles of instruction: Identifying and designing effective, efficient, and engaging instruction. Pfeiffer. 

28. National Academies of Sciences, Engineering, and Medicine, NASEM (2018). Data science for undergraduates: Opportunities and options. Washington, DC: The National Academies Press. https://doi.org/10.17226/25104.  

29. NIST Big Data Public Working Group. (2015). NIST big data interoperability framework: Volume 1, big data definition. https://nvlpubs.nist.gov/nistpubs/SpecialPublications/NIST.SP.1500-1.pdf 

30. Prinsloo, P., \& Slade, S. (2017). Ethics and learning analytics: Charting the (un) charted. SoLAR. 

31. Salinas-Navarro, D. E., Vilalta-Perdomo, E., Michel-Villarreal, R., \& Montesinos, L. (2024). Using generative artificial intelligence tools to explain and enhance experiential learning for authentic assessment. Education Sciences, 14(1), 83. 

32. Sottilare, R., Brawner, K., Goldberg, B., \& Holden, H. (2012). The generalized intelligent framework for tutoring (GIFT). US Army Research Laboratory. 

33. Sottilare, R. A., Burke, C. S., Salas, E., Sinatra, A. M., Johnston, J. H., \& Gilbert, S. B. (2018). Designing adaptive instruction for teams: A meta-analysis. International Journal of Artificial Intelligence in Education, 28(2), 225-264. 

34. Spector, J. M. (2005). Spector's Educratic oath: Innovations in instructional technology. In J. M. Spector, C. Ohrazda, A. Van Schaack, \& D. A. Wiley (Eds.), Innovations in instructional technology: Essays in honor of M. David Merrill (pp. xxxi-xxxvi). Erlbaum. 

35. Siemens, G., \& Gašević, D. (2012). Special issue on learning and knowledge analytics. Educational Technology \& Society, 15(3), 1–163. 

36. Siemens, G. (2013). Learning analytics: The emergence of a discipline. American Behavioral Scientist, 57(10), 1380-1400. 

37. Tomlinson, C. A. (2001). How to differentiate instruction in mixed-ability classrooms. Ascd. 

38. Tsai, Y. S., Perrotta, C., \& Gašević, D. (2020). Empowering learners with personalised learning approaches? Agency, equity and transparency in the context of learning analytics. Assessment \& Evaluation in Higher Education, 45(4), 554-567. 

39. Vandewaetere, M., Desmet, P., \& Clarebout, G. (2011). The contribution of learner characteristics in the development of computer-based adaptive learning environments. Computers in Human Behavior, 27(1), 118-130. 

40. Vygotsky, L. S. (1978). Mind in society: The development of higher psychological processes (Vol. 86). Harvard university press. 

41. Liang Zhang, Jionghao Lin, Ziyi Kuang, Sheng Xu, Xiangen Hu, 2024, SPL: A Socratic Playground for Learning Powered by Large Language Model, https://arxiv.org/abs/2406.13919.  

42. Zimmerman, B. J., \& Schunk, D. H. (2011). Self-regulated learning and performance: An introduction and an overview. Handbook of self-regulation of learning and performance, 15-26. 

43. Zhou, M., \& Winne, P. H. (2012). Modeling academic achievement by self-reported versus traced goal orientation. Learning and Instruction, 22(6), 413–419. https://doi.org/10.1016/j.learninstruc.2012.03 .  
 
\end{document}